\documentclass[twocolumn]{aastex63}
\usepackage{bm}

\newcommand{\mytilde}{\raise.19ex\hbox{$\scriptstyle\sim$}}

\usepackage[utf8]{inputenc}
\usepackage{CJK}
\usepackage{amsmath}
\usepackage{lineno}

\received{}
\revised{}
\accepted{}

\submitjournal{ApJ}

\shorttitle{Precise WL+SL Mass Reconstruction of A2744}
\shortauthors{Cha et al. }

\begin{document}
\title{Precision {\tt MARS} Mass Reconstruction of Abell 2744: \\ Synergizing the Largest Strong Lensing and Densest Weak Lensing Datasets from JWST}

\correspondingauthor{M. James Jee}
\email{sang6199@yonsei.ac.kr, mkjee@yonsei.ac.kr}

\author[0000-0001-7148-6915]{Sangjun Cha}
\affiliation{Department of Astronomy, Yonsei University, 50 Yonsei-ro, Seoul 03722, Korea}
\author[0000-0002-2550-5545]{Kim HyeongHan}
\affiliation{Department of Astronomy, Yonsei University, 50 Yonsei-ro, Seoul 03722, Korea}
\author[0009-0009-4086-7665]{Zachary P. Scofield}
\affiliation{Department of Astronomy, Yonsei University, 50 Yonsei-ro, Seoul 03722, Korea}
\author[0000-0001-9139-5455]{Hyungjin Joo}
\affiliation{Department of Astronomy, Yonsei University, 50 Yonsei-ro, Seoul 03722, Korea}
\author[0000-0002-5751-3697]{M. James Jee}
\affiliation{Department of Astronomy, Yonsei University, 50 Yonsei-ro, Seoul 03722, Korea}
\affiliation{Department of Physics and Astronomy, University of California, Davis, One Shields Avenue, Davis, CA 95616, USA}

\begin{abstract}
We present a new high-resolution free-form mass model of Abell 2744,
combining both weak-lensing (WL) and strong-lensing (SL) datasets from JWST. The SL dataset comprises 286 multiple images, presenting the most extensive SL constraint to date for a single cluster. The WL dataset, employing photo-$z$ selection, yields a source density of $\mytilde350~\rm arcmin^{-2}$, marking the densest WL constraint ever.
The combined mass reconstruction enables the highest-resolution mass map of Abell 2744 within the $\mytilde1.8~$Mpc$\times$1.8~Mpc reconstruction region to date, revealing an isosceles triangular structure with two legs of $\mytilde1$~Mpc and a base of $\mytilde0.6$~Mpc. Although our algorithm MAximum-entropy ReconStruction ({\tt MARS}) is entirely blind to the cluster galaxy distribution, the resulting mass reconstruction remarkably well traces the brightest cluster galaxies with the five strongest mass peaks coinciding with the five most luminous cluster galaxies within $\lesssim 2\arcsec$. We do not detect any unusual mass peaks that are not traced by the cluster galaxies, unlike the findings in previous studies. 
Our mass model shows the smallest scatters of SL multiple images in both source ($\mytilde0\farcs05$) and image ($\mytilde0\farcs1$) planes, which are lower than the previous studies by a factor of $\mytilde4$.
Although {\tt MARS} represents the mass field with an extremely large number of $\mytilde$300,000 free parameters, it converges to a solution within a few hours thanks to our utilization of the deep learning technique.
We make our mass and magnification maps publicly available.
\end{abstract}

\section{Introduction} 
\label{sec:intro}

Gravitational lensing is the most powerful method for reconstructing the mass distribution of astronomical objects without any dynamical assumptions. Cluster mass reconstruction based on lensing is typically classified into two types: weak lensing (WL) and strong lensing (SL). WL uses shape distortions of background galaxies \citep[e.g.,][]{1993ApJ...404..441K, 2000ApJ...532...88H, 2006ApJ...642..720J, 2011ApJ...729..127U, 2017ApJ...851...46F, 2021MNRAS.505.3923S, 2021ApJ...923..101K, 2023ApJ...953..102F}, while SL employs multiply-lensed images \citep[e.g.,][]{2005ApJ...621...53B, 2006ApJ...640..639Z, 2014MNRAS.444..268R, 2015MNRAS.452.1437J, 2016ApJ...819..114K, 2016A&A...587A..80C, 2016MNRAS.459.3447D, 2016A&A...588A..99L, 2018MNRAS.473..663M, 2019MNRAS.485.3738L, 2021ApJ...919...54Z, 2021A&A...646A..57V, 2023A&A...670A..60B}. 

WL can provide measurements for a wide area, sometimes extending beyond the virial radius of galaxy clusters. Its limitation is the intrinsic shape noise/alignment of background galaxies and systematics from instruments. In general, WL requires averaging of background galaxy shapes to reduce the shape noise, which inevitably smooths out substructures. The quality and resolution of WL mass reconstruction depend on the density of the source galaxies and the reconstruction algorithms \citep{bartelmann2001, Schneider2006}.

SL enables us to obtain more precise and higher-resolution mass maps. Since SL uses the positions of individual multiply-lensed features, it is immune to intrinsic source shapes and instrumental systematics, providing better performance in S/N and resolution.
However, since SL features are observed only in the vicinity of cluster centers, SL mass reconstruction techniques alone cannot provide direct constraints beyond this so-called SL regime. SL mass reconstruction techniques quality improves with the number of multiple images and their redshift information availability.

To attain a cohesive understanding of the cluster mass profile from the central region to the outskirts, one of the most effective approaches is to combine the WL and SL signals \citep[e.g.,][]{2005A&A...437...39B, 2006A&A...458..349C, 2007ApJ...661..728J, 2009A&A...500..681M, 2012MNRAS.420.3213O, 2014MNRAS.440.1899G, 2015ApJ...801...44Z}.
However, mass reconstruction with the optimal combination of WL and SL signals is challenging for various reasons.
For the mass models based on analytic profiles, degrees of freedom are too low to properly account for both SL and WL signals simultaneously and sufficiently with high fidelity (underfitting).
Free-form methods may have advantages in this regard because of their high degrees of freedom. However, free parameters usually outnumber observables, and thus the solutions are not unique unless carefully regularized (overfitting).
Aside from these technical challenges, the scarcity of high-quality datasets providing WL and SL has also been an obstacle. Whereas SL data have been predominantly provided by {\it Hubble Space Telescope} observations, wide-field imaging data for WL measurements have been obtained mostly through ground-based observations.
Thus, the severe disparity in mass resolution between the two regimes makes it difficult to implement an optimal smoothing scheme that preserves significant substructures while suppressing noisy fluctuations.

In this study, we present a new high-resolution free-form mass model of Abell 2744 (A2744 hereafter) at $z=0.308$ from the central region to the outskirts by combining unprecedentedly large WL and SL datasets, utilizing the recent high-quality wide-field ($\mytilde7 \farcm 6\times 7 \farcm 6$) JWST images. The SL dataset consists of 286 multiple images, whereas the WL source density based on photo-z selection reaches $\mytilde350$~arcmin$^{-2}$, which is the largest dataset ever used for cluster mass reconstruction.

The galaxy cluster A2744, nicknamed Pandora's cluster, was introduced as a peculiar cluster due to its complex and puzzling substructures by \citet{2011MNRAS.417..333M}.
There is an offset between its intracluster medium (ICM) and mass~\citep{2010MNRAS.406.1134S}, resembling the ``Bullet Cluster" \citep{2006ApJ...648L.109C}. Also, \citet{2011MNRAS.417..333M} reported the presence of a mass clump without a counterpart in luminous galaxies\footnote{The authors referred to the feature as ``ghost."}.
The cluster was one of the prime targets in the Hubble Frontier Fields (HFF) program \citep{2015ApJ...800...84C, 2017ApJ...837...97L} and has been a subject of a number of studies encompassing lensing
\citep[e.g.,][]{2015ApJ...811...29W, 2016MNRAS.463.3876J, 2016ApJ...817...24M, 2019MNRAS.488.3251S, 2022AAS...24021405F, 2023ApJ...951..140C}, X-ray \citep[e.g.,][]{2015Natur.528..105E}, and  radio observations \citep[e.g.,][]{2013A&A...551A..24V, 2017ApJ...845...81P, 2021A&A...654A..41R}.

Recently, under the program Ultra-deep NIRCam and NIRSpec ObserVations before the Epoch of Reionization (UNCOVER) \citep[][]{2022arXiv221204026B, 2023arXiv230102671W}, deep wide-field ($\mytilde45$~arcmin$^2$) JWST imaging observations were carried out. Utilizing the unprecedented depth and resolution of the new A2744 JWST data, several interesting scientific results have been published \citep[e.g.,][]{2023ApJ...952..142F, 2023ApJ...947L..24M, 2023ApJ...948L..14C, 2023ApJ...948L..15V, 2023MNRAS.524.5486A}. 
However, in the case of the lens model, only lensing models based on parametric SL techniques have been presented. They employ the light-trace-mass (LTM) assumption to model the masses of the cluster galaxies. The center positions of the cluster-scale dark matter halos are allowed to move within the prior intervals centered on the BCGs, with $\lesssim 3 \arcsec$ and $\lesssim 5\arcsec-30 \arcsec$ for \citet{2023MNRAS.523.4568F} and  \citet{2023ApJ...952...84B}, respectively.

This study provides the first non-LTM and profile-independent mass reconstruction using not only multiple images (SL) but also distortion (WL) signals, which are densely distributed across the entire A2744 JWST field. We employ the MAximum-entropy ReconStruction ({\tt MARS}) algorithm \citep{2022ApJ...931..127C, 2023ApJ...951..140C}. {\tt MARS} is a free-form mass reconstruction method that utilizes cross-entropy to regularize its solution, resulting in a quasi-unique solution despite the number of free parameters greatly exceeding that of the observables. 
\citet{2022ApJ...931..127C} tested the fidelity of {\tt MARS} with the synthetic  cluster data \citep{2017MNRAS.472.3177M} and demonstrated that {\tt MARS} is one of the best-peforming methods. The test with real HFF cluster data showed that the image-plane scatter of the multiple images is lower than any of the publicly available mass models \citep{2023ApJ...951..140C}.
In the current study, we extend the previous {\tt MARS} algorithm of \citet{2022ApJ...931..127C} to accommodate WL constraints as well.

This paper is structured as follows. In \textsection\ref{sec:data}, we introduce the JWST NIRCam imaging data and reduction steps. In \textsection\ref{sec:method}, we describe the WL analysis method and our algorithm for the mass reconstruction combining WL and SL. We show our results in \textsection\ref{sec:result}. In \textsection\ref{sec:discuss}, we discuss our results and conclude in \textsection\ref{sec:conclusion}.
Unless stated otherwise, we assume a flat $\Lambda$CDM cosmology with the dimensionless Hubble constant parameter $h=0.7$ and the matter density $\Omega_{M}=1-\Omega_{\Lambda}=0.3$. The plate scale at the cluster redshift ($z=0.308$) is $4.536 ~\rm kpc ~\rm arcsec^{-1}$.

\section{Data} \label{sec:data}
\subsection{JWST NIRCam Images}
Our WL+SL analysis utilizes the publicly available NIRcam mosaic images of A2744 processed by the UNCOVER team\footnote{\url{https://jwst-uncover.github.io/DR1.html}}, who combined the three JWST programs: 1) JWST-DD-ERS-1324 \citep[PI: T. Treu;][]{2022ApJ...935..110T}, 2) JWST-GO-2561 \citep[PIs: I. Labbe and R. Bezanson;][]{2022arXiv221204026B}, and 3) JWST-DD-2756 \citep[PI: W. Chen;][]{2022TNSAN.257....1C}. 
We created color-composite images from the 7 filter imaging data (F115W, F150W, F200W, F277W, F356W, F410M, and F444W) and identified additional multiple images used for SL constraints, based on their morphological properties and colors.
We use F200W to measure WL because it provides the optimal scale for the point spread function (PSF) sampling \citep{2023ApJ...953..102F}.
In order to obtain the PSF model on the mosaic images, it is necessary to derive PSF models for input frames and stack them with proper rotations and dithers. Thus, we retrieved the input {\tt CAL} files from the Mikulski Archive for Space Telescopes (MAST)\footnote{\url{https://archive.stsci.edu/}} and found the coordinate transformation from the detector reference frame to the mosaic reference frame for each input {\tt CAL} image.
For more details on the mosaic image, we refer readers to \citet{2023arXiv230102671W}.

\subsection{SL Data}\label{SL_data_section}

\begin{figure*}
\centering
\includegraphics[width=\textwidth]{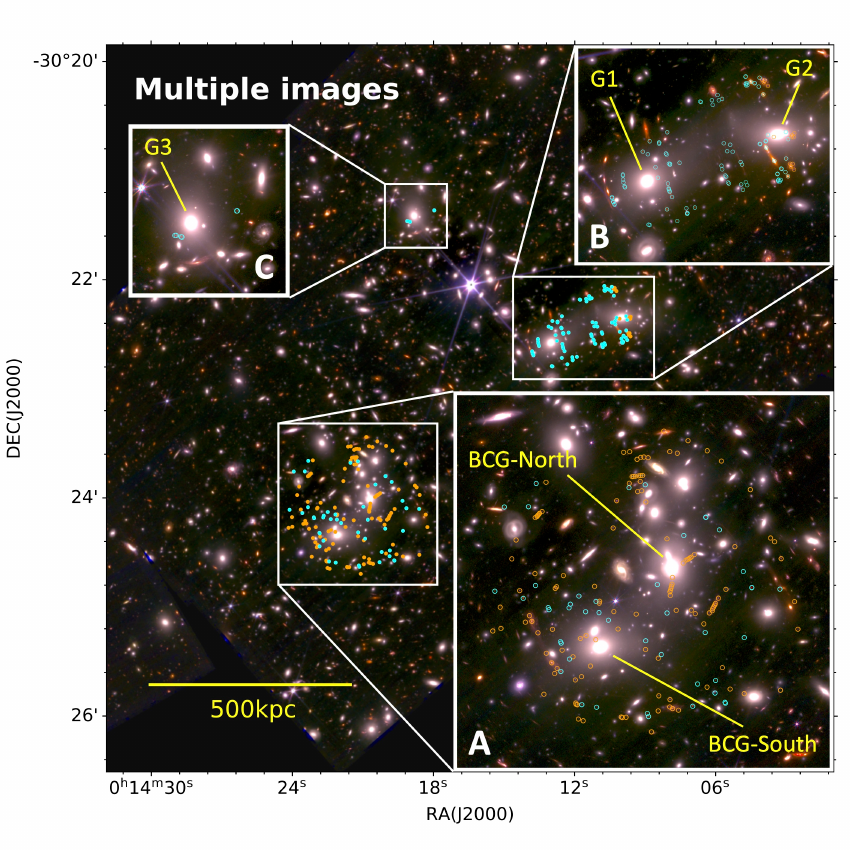} 
\caption{Multiple image distributions in the A2744 field. Orange (cyan) circles indicate the locations of the gold-class (silver-class) multiple images, including the knots of extended multiple images. The color-composite image is created using the F444W filter for red, F277W filter for green, and F115W filter for blue. The notation from \citet{2023ApJ...952...84B} is followed for labeling the 5 brightest galaxies. The displayed field of view is $400\arcsec\times400\arcsec$.}
\label{SL_catalog}
\end{figure*}

\begin{figure*}
\centering
\includegraphics[width=0.9\textwidth]{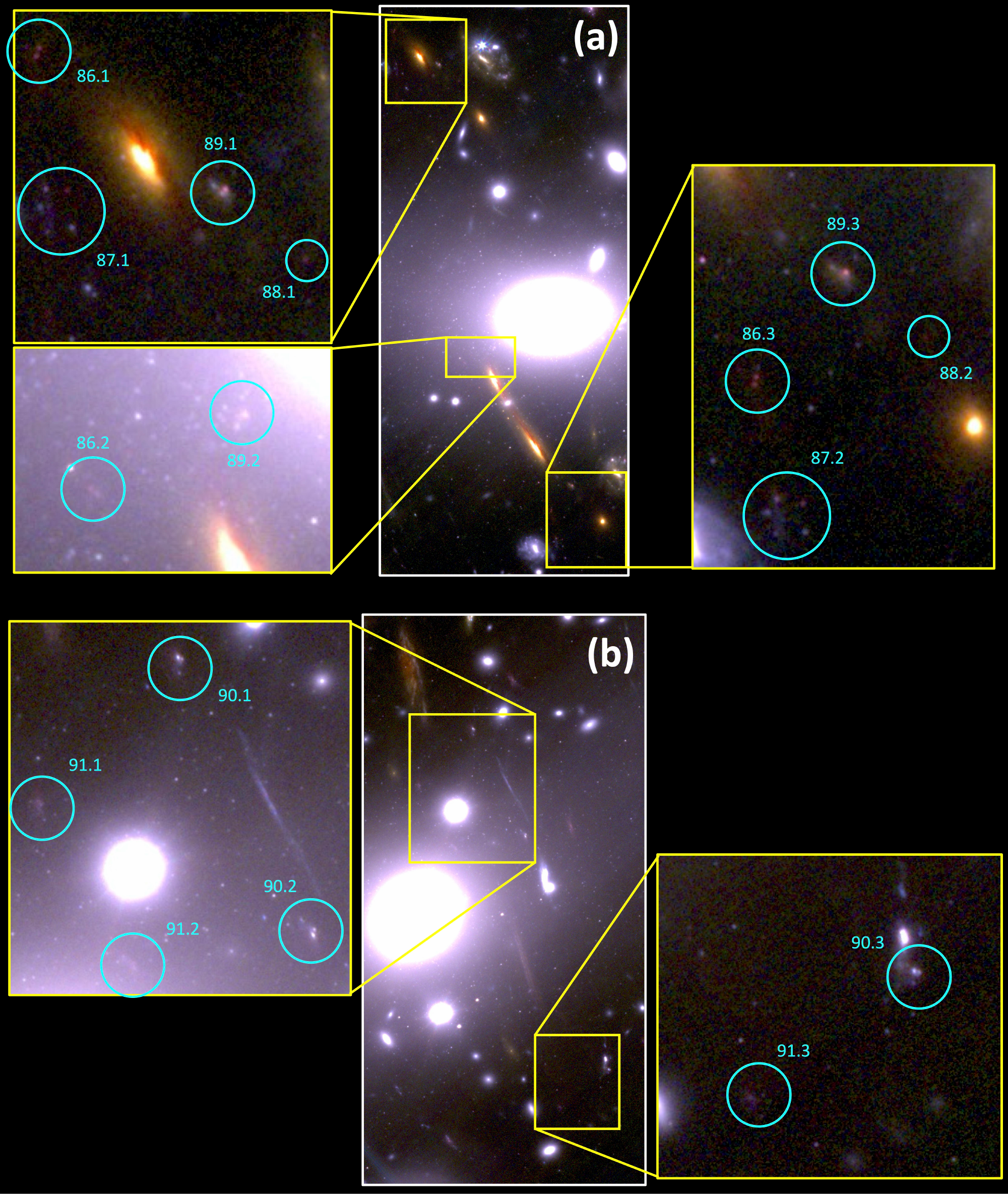} 
\caption{Newly identified multiple image candidates from this study. We find 6 candidate SL systems around G2 (a) and G1 (b); see Figure~\ref{SL_catalog} for their locations. The cyan circles and numbers indicate the locations and IDs of the multiple images, respectively. The color-composite images are created using F200W (red), F150W (green), and F115W (blue).}
\label{newly_identified}
\end{figure*}

\begin{deluxetable}{cccccc}\label{multiple_catalog_a2744}
\tablecaption{Multiple image catalog of A2744.}
\tablehead {
\colhead{ID} &
\colhead{R.A. (J2000)} &
\colhead{Decl. (J2000)} &
\colhead{z$^1$} & 
\colhead{Class$^2$}
}
\startdata
1.1	&	39.967047	&	-1.5769172	&	0.8041	&	gold    \\
1.2	&	39.976273	&	-1.5760558	&	0.8041	&	gold	\\
1.3	&	39.968691	&	-1.5766113	&	0.8041	&	gold	\\
2.1	&	39.973825	&	-1.584229	&	0.7251	&	gold	\\
2.2	&	39.971003	&	-1.5850422	&	0.7251	&	gold	\\
$\cdots$	&		&		&		&		&		\\
\enddata
\tablecomments{$^1$Redshifts indicated with an asterisk (*) represent the redshifts estimated using {\tt MARS} (see \textsection\ref{reconstruction_method} for details). 
$^2$``Class" is described in \textsection\ref{SL_data_section}. 
$^3$The complete multiple-image catalog is available as online supplementary material.}
\end{deluxetable}

We combine SL data from various sources in the literature and identify new multiple images.
We classify the multiple images into gold and silver classes. Gold-class images are those with spectroscopic redshifts. Silver-class images do not possess spectroscopic redshifts, but they either have photometric redshifts or have been identified as multiple images by various studies \citep[e.g.,][]{2015MNRAS.452.1437J, 2016ApJ...819..114K, 2018MNRAS.473..663M, 2023A&A...670A..60B, 2023MNRAS.523.4568F, 2023ApJ...952...84B}.

Figure~\ref{SL_catalog} shows the multiple image distributions that we compiled for the current study.
In region A, we adopt the catalog of \citet{2023ApJ...951..140C}, who compiled the SL data from \cite{2015MNRAS.452.1437J}, \citet{2016ApJ...819..114K}, \citet{2018MNRAS.473..663M}, and \citet{2023A&A...670A..60B}.
All multiple images in \citet{2023A&A...670A..60B} have spectroscopic redshifts. If a multiple-image system without a spectroscopic redshift is agreed to be a valid system by the three papers \citep{2015MNRAS.452.1437J,2016ApJ...819..114K, 2018MNRAS.473..663M}, the system is classified as a silver class.
In regions B and C, we adopt 79 multiple images from \citet{2023MNRAS.523.4568F} and 9 multiple images, which are knots (distinctive features such as star-forming regions in the extended multiple images) of system 68 from \citet{2023ApJ...952...84B}. 
If neither spectroscopic nor photometric redshift information is available, or if there is a considerable disagreement among the photometric redshifts within the same system
(greater than $0.15\Delta z / (1 + \overline{z_{phot}})$, where $\overline{z_{phot}}$ indicates the mean of the photometric redshifts within the same system), we treat the redshift of the system as a free parameter. 
We also free the redshift when {\tt MARS} cannot converge multiple images with its input photometric redshift
(see \textsection\ref{reconstruction_method} for details).

In addition to the compiled catalog above, we have identified 16 new multiple-image candidates 
from 6 systems
in region B (Figure~\ref{newly_identified}) 
and all of them are classified as silver-class images.
When multiple photometric redshifts are available for a system, we adopt the mean value as the system redshift.

A total of 286 images (136 gold and 150 silver images), including 91 knots, are utilized for mass reconstruction of A2744. We list multiple images in Table~\ref{multiple_catalog_a2744}, where we follow the numbering scheme of \citet{2023MNRAS.523.4568F} for the multiple images in regions B and C.

\subsection{WL Data}
\subsubsection{PSF Modeling}\label{psf_modeling}
\begin{figure*}
\centering
\includegraphics[width=0.45\textwidth]{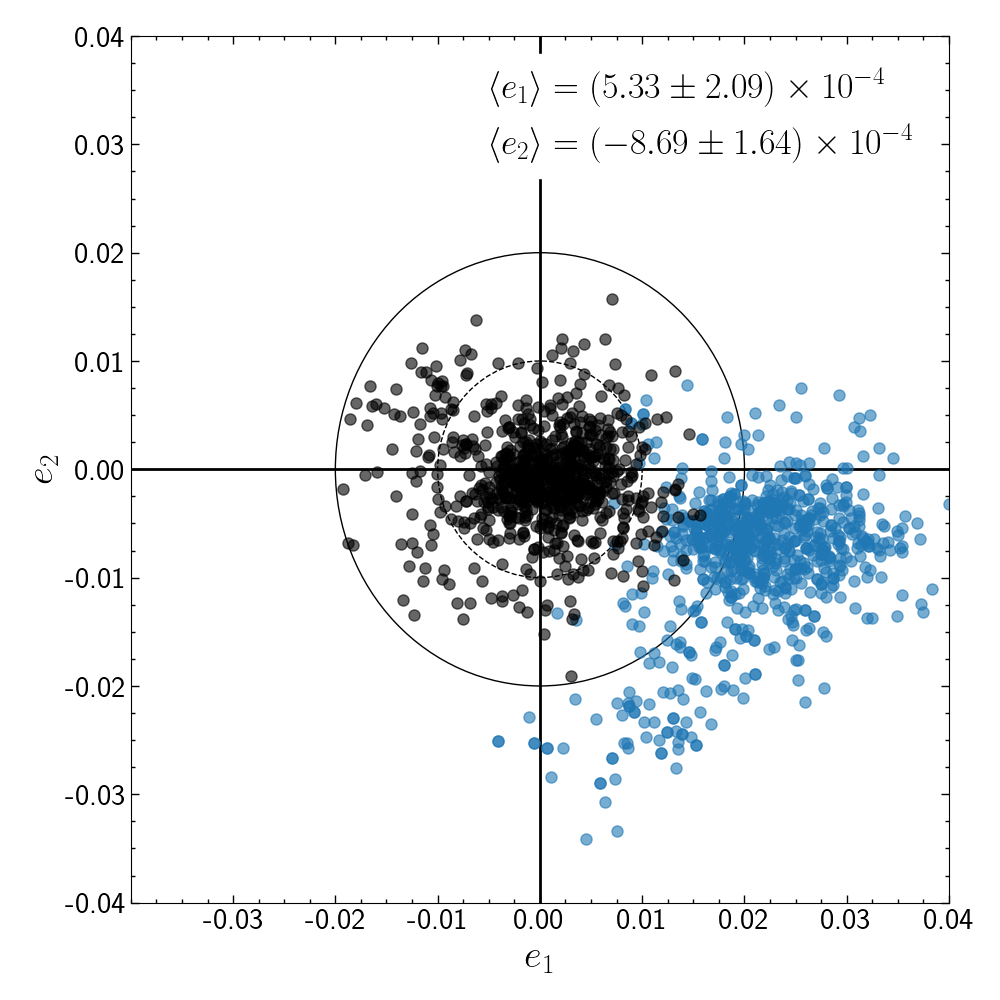} 
\includegraphics[width=0.45\textwidth]{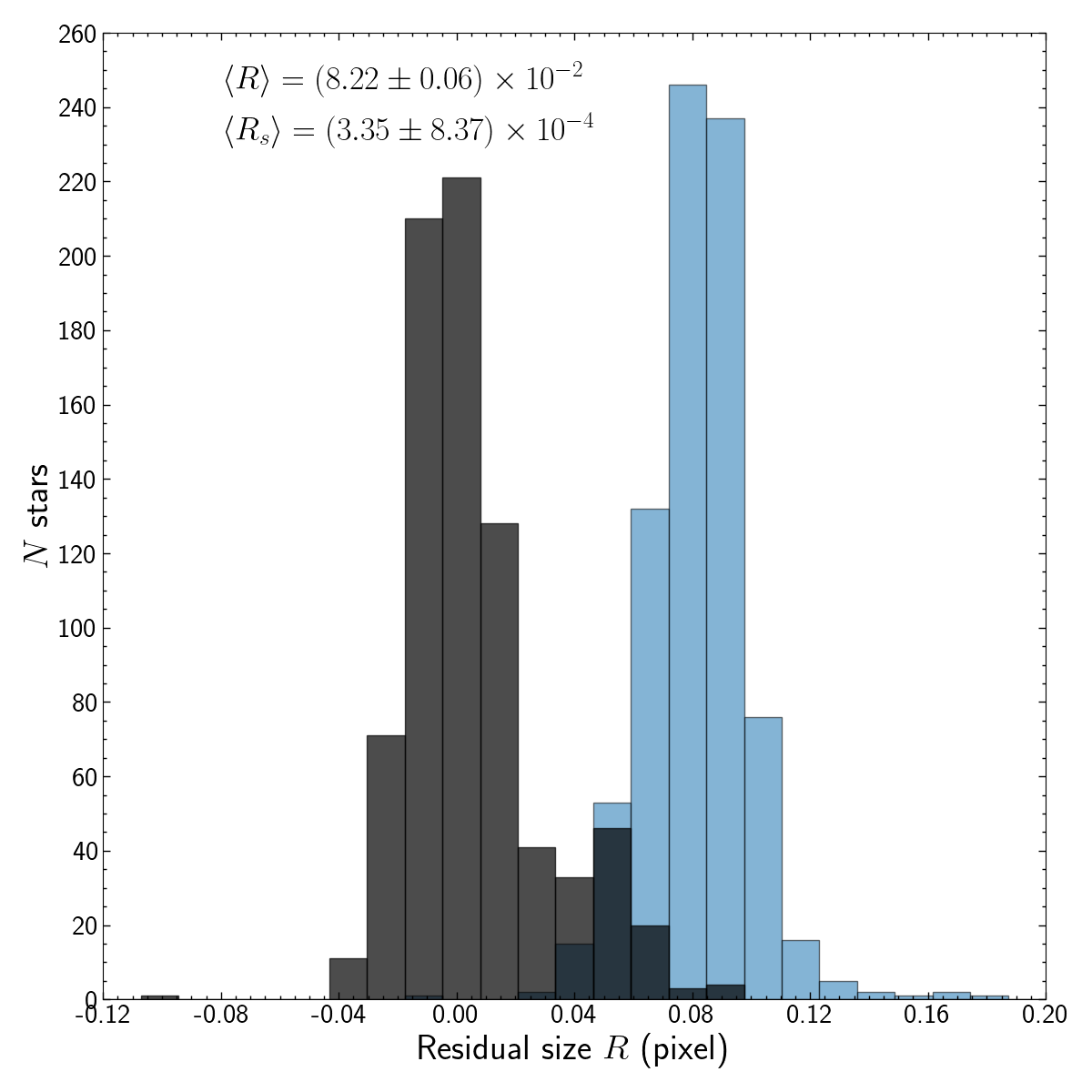} 
\caption{\texttt{WebbPSF} model corrections for ellipticity and size. We present the result for F200W, where we measure WL signals. Left: Original complex ellipticity components of stars measured in the A2744 NIRCam images (blue) and residuals (black) computed by subtracting the \texttt{WebbPSF} model values from the observed values. The median and standard error for the residuals are provided in the top right corner of the plot. Right: The residual size $R$ between the observed stars and the PSFs (blue) and the residual computed after applying a difference kernel (black). In the top left  of the plot, the median residuals before 
$\left<R\right>$ and after the kernel $\left<R_s\right>$, including the associated standard errors, are provided.}
\label{webbpsf_plots}
\end{figure*}

In order to accurately perform WL analysis, one must model and correct for the PSF. The relatively small field of view of NIRCam provides only a small number of stars per pointing. Thus, in some cases producing an empirical PSF model across the detector based on the limited number of observed stars is not feasible. \citet{2007PASP..119.1403J} overcame this problem for HST by building a PSF library from the observations of dense stellar fields such as globular clusters and then using the observed stars in a WL field to find the matching library.
However, currently, the amount of NIRCam data for dense stellar fields is not sufficient to build such an extensive PSF library as in \citet{2007PASP..119.1403J}.
Instead, in the current study, we employ the wavefront sensor data of JWST, which provide optical path difference (OPD) maps. To utilize these OPD maps, we use the package \texttt{WebbPSF} \citep{10.1117/12.925230, 10.1117/12.2056689}. 

\texttt{WebbPSF} provides two OPD maps: one from before and the other from after the input (observation) date. The OPD map closest to the date of observation cannot always be guaranteed to be the best choice since JWST's optical alignments are monitored and adjusted on a regular basis. For the A2744 data, we
verified that both OPD maps give similar results when compared with the PSF in the input frame. Therefore, we decided to choose the OPD map closest to each observation date to maintain consistency.   

To verify that the PSFs are properly reconstructed by \texttt{WebbPSF}, we first collect star postage-stamp images for each detector from the A2744 data. Then, the PSF reconstruction for each star is produced using the selected OPD map for the given observation date. These star stamps are taken from calibrated, non-distortion corrected images with the native pixel scale of the NIRCam short wavelength channel. The residuals between the stars and the model are computed for the size and ellipticity. These shape parameters are measured using the following quadrupole moments \citep{2007PASP..119.1403J}:
\begin{equation}
    Q_{i,j} = \frac{\int d^2 \theta W(\theta)I(\theta)(\theta_i - \bar{\theta}_i)(\theta_j - \bar{\theta}_j)}{\int d^2 \theta W(\theta)I(\theta)} \;, \quad i,j \in {1,2} \;.
\end{equation}
$I(\theta)$ is the pixel intensity at $\theta$, $\bar{\theta}_{i,j}$ is the center of the star, and $W(\theta)$ is a circular Gaussian weight function utilized to suppress noise in the outer regions of the PSF. We can then define the complex ellipticity components ($e_1$, $e_2$) and size ($R$):
\begin{align}
    e_1 + i e_2 & = \frac{Q_{11} - Q_{22} + 2iQ_{12}}{Q_{11}+Q_{22}+2(Q_{11}Q_{22}-Q^2_{12})^{\frac{1}{2}}} \;, \\ 
    R & = \sqrt{Q_{11} + Q_{22}} \;.
\end{align}
For an ellipse with semi-major and -minor axes $a$ and $b$ and position angle $\phi$ (measured counterclockwise from the reference axis), $e_1$ and $e_2$ correspond to
\begin{equation}
    e_1 = \frac{a-b}{a+b} \cos(2\phi) \;, \quad e_2 = \frac{a-b}{a+b} \sin(2\phi)
\end{equation}
with $(a-b)/(a+b)$ referred to as the ellipticity $e$.

The left panel of Figure \ref{webbpsf_plots} shows the ellipticities of the observed stars and the residuals (observed star ellipticity $-$ model PSF ellipticity). The detector-induced ellipticity in the observed stars is corrected by the model, as the residual points are centered around $e_1, e_2 = 0$. The right panel shows the residual size $R$ computed by subtracting the observed model size from the star size. The initial result is given as the blue distribution, with the median being $\left< R \right> = (8.22 \pm 0.06) \times 10^{-2}$. The PSFs produced by the \texttt{WebbPSF} package are systematically smaller than the observed stars, and this difference is likely due to \texttt{WebbPSF} not sufficiently taking into account detector effects such as interpixel capacitance or intrapixel sensitivity variations. To remedy this issue, a difference kernel is applied to the PSFs; we used a Gaussian kernel with an empirically determined kernel size. Represented by the black histogram, the residuals after applying the kernel are approximately centered at $R = 0$, with the median being $\left<R_s\right> = (3.35 \pm 8.37) \times 10^{-4}$. Given the small magnitudes for both the ellipticity and size residuals ($\sim10^{-4}$), our PSF model produced by \texttt{WebbPSF} and the difference kernel should be sufficient for WL analysis.

After ensuring that the chosen model can effectively reproduce the detector effects seen in NIRCam data, a distortion-corrected and oversampled PSF model is required for use with the A2744 mosaic image. For each of the eight NIRCam detectors (NRCA1-4 and NRCB1-4) and for each observation date, one hundred evenly-spaced (distortion corrected and resampled with a pixel scale of 0.02) 31~pixel $\times$31~pixel PSFs are created using \texttt{WebbPSF}. A principal component analysis (PCA) is then performed to allow for the production of a 31~pixel $\times$31~pixel PSF at any detector position. Although \texttt{WebbPSF} could be used to produce a PSF at each galaxy centroid position, the high source count and the potential need for multiple stacked and rotated PSFs per galaxy make PCA a more practical choice given the lengthy execution time required by \texttt{WebbPSF}. A more detailed description of PSF modeling with PCA can be seen in \citet{2007PASP..119.1403J}. To model a PSF at each galaxy position in the mosaic image, we identify all contributing input frames, retrieve their PSF models, and stack the results with proper weights and rotations.

The distribution of stars across the A2744 field allows for PSF modeling with PCA using the mosaic image in the same manner as in \citet{2023ApJ...953..102F}, which can be compared to the modeling method used in this study. Further analysis and justification for the choice of PSF modeling strategy can be found in Appendix~\ref{dif_psf_results}.

\subsubsection{Ellipticity Measurement \& Source Selection}\label{shape_measure}
We use a forward-modeling approach to measure the ellipticity of each galaxy before it is convolved by the JWST PSF.
The PSF model predicted at the location of each source was convolved with an elliptical Gaussian and fit to the galaxy using the {\tt MPFIT} optimizer \citep{2009ASPC..411..251M}.
We fixed the background and centroid of a source to the values output by {\tt SExtractor} \citep{1996A&AS..117..393B}.
The free parameters are the position angle ($\phi$), semi-major and -minor axes ($a$ and $b$), and normalization.
Ideally, the average of the galaxy ellipticity $(e_1,e_2)=(e \cos 2\phi, e\sin 2\phi)$ should be an unbiased estimator of the reduced shear $g$. However, because of a number of factors, in practice, the estimator is biased. Among them, two outstanding contributors are ``noise bias" and ``model bias." Noise bias occurs due to the nonlinear relation between pixel noise and parameter noise whereas model bias is caused by the fact that the galaxy model (in this case, the elliptical Gaussian) is different from the actual galaxy profile. In addition, the blending effect is also a significant source of shear bias.
Instead of characterizing these biases individually, we performed WL image simulations matching the JWST quality and derived multiplicative factors of 1.11 and 1.07 for $g_1$ and $g_2$, respectively \citep[][]{2013ApJ...765...74J, 2023ApJ...953..102F}.

To select source galaxies, we utilized the photometric redshift catalog provided by the UNCOVER team\footnote{\url{https://jwst-uncover.github.io/DR1.html}}.
As a conservative measure, we selected sources with a photometric redshift greater than 0.4 as background objects.
Additionally, we imposed shape quality criteria based on fitting status and recovered shapes. We discarded sources whose {\tt MPFIT STATUS} parameter is different from unity because this typically indicates unstable fitting. Also, the minimum ellipticity measurement error is set to $\delta e=0.4$. When the object size is reported to be too small, the source is typically either point source-like or unrealistically compact. We avoided these cases by imposing that the semi-minor axis is greater than 0.4 pixels. 
The mean photometric redshift of the sources is 2.5, and the source density is $\sim350~\rm arcmin^{-2}$, which is the highest among all existing WL studies (the typical source density in HST-based WL is $\mytilde100~\rm arcmin^{-2}$).

We note that we did not explicitly mask out the SL areas when selecting WL sources. In general, because galaxy shapes become curved in the SL regime, their ellipticities may underrepresent the local reduced shear.
We find that approximately 60 objects are located within $0\farcs2$ of the multiple images. Since they comprise only $\mytilde0.6$\% of the WL sources, we do not think that the bias caused by these sources is significant. Furthermore, given the much stronger constraining power from the SL multiple images in the SL regime, the bias, if any, should be negligible.

\section{Method} \label{sec:method}

\subsection{Lensing Theory}
In this section, we provide a brief review of lensing theory, covering the range from the outskirts (WL) to the central regions (SL) of galaxy clusters. For more details, we refer readers to review papers \citep[e.g.,][]{bartelmann2001, Kochanek2006, 2011A&ARv..19...47K, hoekstra2013}.
In the WL regime, the characteristic scale of the variation in the distortion of the background galaxy image becomes much smaller than the galaxy size, and thus
the change in the galaxy shape is approximated by the following matrix $\mathbf{A}$:
\begin{equation}
    \mathbf{A}= (1-\kappa)
    \begin{pmatrix}
    1-g_1   &   -g_2 \\
    -g_2    &   1 + g_1
    \end{pmatrix}, \label{eqn_A}
\end{equation}
where $\kappa$ indicates the convergence and $g_{1(2)}$ denotes the first (second) component of the reduced shear $g=({g_1}^2+{g_2}^2)^{1/2}$. 
The reduced shear $g$ is computed as $g=\gamma/(1-\kappa)$, where $\gamma$ is shear. 

The convergence $\kappa$ is given by:
\begin{equation}
    \kappa=\frac{\Sigma}{\Sigma_c},
\end{equation}
where $\Sigma$ ($\Sigma_c$) is the (critical) surface mass density. $\Sigma_c$ can be computed as follows:
\begin{equation}
    \Sigma_c = \frac{c^2 D_s}{4 \pi G D_d D_{ds}},
\end{equation}
where $c$ is the speed of light, $D_{s(d)}$ denotes the angular diameter distance to the source (lens), and $D_{ds}$ represents the angular diameter distance between the lens and the source.
The shear $\gamma$ is related to the convergence $\kappa$ through the following:
\begin{equation}
    \bm{\gamma}\mathbf{(x)}=\frac{1}{\pi}\int\mathbf{D}(\mathbf{x}-\mathbf{x'})\kappa(\mathbf{x'})d\mathbf{x'},
    \label{eqn_shear}
\end{equation}
where the kernel $\mathbf{D}$ at the position $(x_1,x_2)$ is defined as:
\begin{equation}
    \mathbf{D}=-\frac{1}{(x_1 - \mathbf{i}x_2)^2}.
\end{equation}

In the SL regime, the absolute value of the reduced shear can exceed unity ($|\bm{g}|>1$). In this case, we replace $\bm{g}$ with $1/\bm{g^*}$ in equation~\ref{eqn_A}, where $\bm{g^*}$ represents the complex conjugate of $\bm{g}$.

The relation between the observed image position $\bm{\theta}$ and the source position $\bm{\beta}$ follows the lens equation:
\begin{equation}
    \bm{\beta}=\bm{\theta}-\bm{\alpha}(\bm{\theta}),
\label{lens_equation}
\end{equation}
where $\bm{\alpha}$ is called the deflection angle. 
The deflection angle $\bm{\alpha}$ can be computed through the convolution of the convergence $\kappa$ or the differentiation of the deflection potential $\Psi$.
The {\tt MARS} algorithm utilizes the convolution to obtain the deflection angle $\bm{\alpha}$ as follows:
\begin{equation}
    \bm{\alpha} (\bm{\theta}) = \frac{1}{\pi} \int
    \kappa (\bm{\theta}^{\prime}) \frac{\bm{\theta}-\bm{\theta}^{\prime}}{|\bm{\theta}-\bm{\theta}^{\prime}|^{2}} \bm{d^{2} {\theta}}^{\prime}. 
    \label{eqn_deflection_via_con}
\end{equation}
Since the mass outside the reconstruction field affects both deflections (equation~\ref{eqn_deflection_via_con}) and shears (equation~\ref{eqn_shear}) within the reconstruction field, we make the field size of the model 40\% larger ($\mytilde2.5~$Mpc$\times2.5$~Mpc) than the actual reconstruction field ($\mytilde1.8~$Mpc$\times1.8$~Mpc).

\subsection{{\tt MARS} WL + SL Mass Reconstruction Algorithm}\label{reconstruction_method}

We employ the {\tt MARS} algorithm \citep[][]{2022ApJ...931..127C, 2023ApJ...951..140C} to reconstruct the mass distribution of A2744. In our previous studies, the application of {\tt MARS} was limited to SL mass modeling. In the current study, we revised {\tt MARS} so that now it can also utilize WL signals.
The new {\tt MARS} minimizes the following function:
\begin{equation}
    f = m\chi^2_{SL} + w\chi^2_{WL} + rR, 
\label{total_eqn}
\end{equation}
where $\chi^2_{SL}$ and $\chi^2_{WL}$ represent the $\chi^2$ terms for SL and WL observables, respectively, and $R$ is the regularization term. The weight parameters $m, w$, and $r$ determine the relative importance of the SL, WL, and regularization terms, respectively.

Reduction of $\chi^2_{SL}$ decreases the scatters of the multiple images in the source plane (i.e., positions of delensed multiple images). $\chi^2_{SL}$ is defined as: 
\begin{equation}
    \chi^{2}_{SL}=\sum_{i=1}^{I} \sum_{j=1}^{J}\frac{(\bm{\theta}_{i,j}-\bm{\alpha}_{i,j}(z)-\bm{\beta}_{i})^{2}}{{\sigma_{i}}^{2}},
\label{chi_square_SL}
\end{equation}
where
\begin{equation}
    \bm{\beta}_{i}=\frac{1}{J}\sum_{j=1}^{J}(\bm{\theta}_{i,j}-\bm{\alpha}_{i,j}(z)).
\end{equation}
$I$ is the total number of systems, and $J$ is the number of multiple images from each system. As is done in \citet{2023ApJ...951..140C}, we treat each ``knot" (distinctive feature such as star-forming region) within a multiple image as an individual image. We refer readers to \citet{2023ApJ...951..140C} for details. 

We define $\chi^2_{WL}$ as follows:
\begin{equation}
    \chi^{2}_{WL}=\sum_{i=1}^{I} \sum_{j=1}^{2}\frac{({g}_{i,j}-\epsilon_{i,j})^2}{\sigma_{s}^2({g}) + \sigma_{m,i}^2},
\label{chi_square_WL}
\end{equation}
where ${g}_{i,j} (\epsilon_{i,j})$ indicates the $j^{th}$ component of the expected reduced shear (observed ellipticity) evaluated at the position and redshift of the $i^{th}$ WL source. $\sigma_{m,i}$ is the measurement error for the $i$th WL source, and $\sigma_s({g})$ is the shape noise per ellipticity component for the expected reduced shear $g$.
In general, the shape noise decreases as $g$ increases (i.e., every source galaxy becomes stretched in a nearly identical way regardless of its intrinsic shape when $g$ approaches unity). We use the following function to model the effect:
\begin{equation}
    \sigma_{s}(g)=(1-{g}^2)\sigma_{s}(0).
\label{intrinsic_shape}
\end{equation}
where $\sigma_{s}(0)=0.25$ is the intrinsic shape noise in the region where there is no shape distortion by gravitational lensing. 

By adopting the maximum cross entropy, {\tt MARS} regularizes the mass reconstruction to prevent overfitting and achieve the smoothest possible solutions unless the substructures are strongly required by the data. The regularization term $R$ is given by: 
\begin{equation}
    R=\sum\left (p-\kappa+\kappa\mathrm{ln} \frac{\kappa}{p} \right),
\label{cross_entropy}
\end{equation}
where $\kappa$ and $p$ are the convergence and prior, respectively. 
The prior is updated for each epoch of minimization by smoothing the convergence map obtained from the previous iteration using a Gaussian kernel. For more details, we refer readers to \citet{2022ApJ...931..127C, 2023ApJ...951..140C}. 
The kernel sizes are $\sigma=0.6, 1.2, 2.4$ pixels for the $140\times140, 280\times280, \rm and ~ 560\times560$ mass grids, respectively. We double the kernel size when only WL data are used.

The WL+SL mass reconstruction run is carried out in the following steps:
\begin{enumerate}
    \item Perform an SL-only (i.e., set $w=0$ in equation~\ref{total_eqn}) low-resolution mass reconstruction with a $140\times140$ mass grid, which includes 20-cell thick stripes at the boundaries outside the reconstruction field (the actual mass reconstruction field has a resolution of $100\times100$). We start minimization with a flat $\kappa=0.1$ convergence field. The minimization ends when the multiple image positions converge in the source plane.
    \item Add the $\chi^2_{WL}$ term to evolve the SL-only solution from step 1 to the WL + SL one.
    \item Increase the resolution of the grid from step 2 to $280\times280$ (by a factor of two) and restart the WL + SL mass reconstruction.
    \item Repeat step 3 by further increasing the resolution to $560\times560$ (now the marginal stripe is 80-cell thick, and the resolution within the reconstruction field is $400\times400$). We end the reconstruction process when the $\chi^2_{WL}$ per WL component reaches unity while the multiple image scatters in the source plane are consistent with noise.
\end{enumerate}

We treat the redshift of a multiple-image system as a free parameter if neither its spectroscopic nor photometric redshift is known. In case {\tt MARS} cannot converge a multiple-image system with its known photometric redshift, we also free its redshift. These freed redshifts are constrained along with the mass reconstruction, and we refer to the values as model redshifts. As done in \citet{2023ApJ...951..140C}, we set a flat prior for a model redshift with $z_{model}=[z_{cluster}+0.1,15]$, where $z_{cluster}=0.308$ is the redshift of A2744. 

Including the peripheral grid cells, the final resolution of the resulting mass map is $560\times560$, requiring $\sim300,000$ free parameters. To minimize our target function $f$ (Equation~\ref{total_eqn}) with this large number of free parameters, we utilize the Adam (adaptive moment) optimizer~\citep{adam2014}. The Adam optimizer is commonly used in deep learning for optimizing complex models with an extremely large number of free parameters. 
Thanks to its efficiency, {\tt MARS} converges to a solution within just a few hours.

Although our main result is the one constrained by both WL and SL data, it is instructive to examine how the solution changes with respect to the full WL+SL reconstruction when only one of the two datasets is used. Therefore, we repeat the above mass reconstruction run with one dataset at a time [for the WL-only (SL-only) reconstruction, we set $m=0 ~ (w=0)$ in equation~\ref{total_eqn}] and include the comparisons in our discussion.

Hereafter, when we present our $\kappa$ field, unless stated otherwise, $\kappa$ is scaled to $D_{ds}/D_{s}=1$. The corresponding critical surface mass density (i.e., $\kappa=1$) is $1.777\times10^9 M_{\odot} ~\rm kpc^{-2}$.

\section{Result} \label{sec:result}
\subsection{Projected Mass Distribution}\label{project_mass}
\begin{figure*}
\centering
\includegraphics[width=0.74\textwidth]{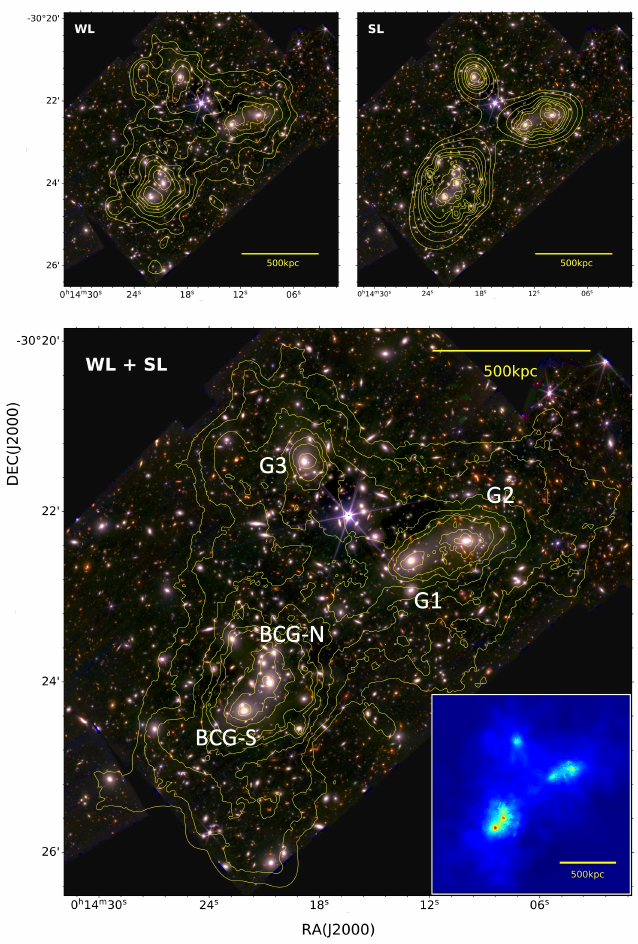} 
\caption{Mass contours of A2744 overlaid on the color-composite images. The yellow contours indicate the convergence $\kappa$. The upper left (upper right) panel displays the mass contours obtained from the WL-only (SL-only) mass map, while the lower panel presents the mass contours derived from the WL + SL mass map. In the WL-only mass map, the contours correspond to $\kappa=[0.15, 0.2, 0.3, 0.4, 0.5, 0.6, 0.7, 0.8]$. For the SL-only and WL + SL mass maps, the contours indicate $\kappa=[0.15, 0.2, 0.3, 0.4, 0.6, 0.9, 1.2, 1.8, 2.4]$. To mitigate pixel-scale artifacts, we apply Gaussian smoothing with a kernel of $\sigma=2\arcsec ~ (\sigma=1\arcsec)$ to the WL-only (WL + SL) mass contours. In the lower right panel, unsmoothed mass distributions are displayed as a color map. The color-composite images are the same as shown in Figure~\ref{SL_catalog}.}
\label{all_massmaps}
\end{figure*}

\begin{figure*}
\centering
\includegraphics[width=\textwidth]{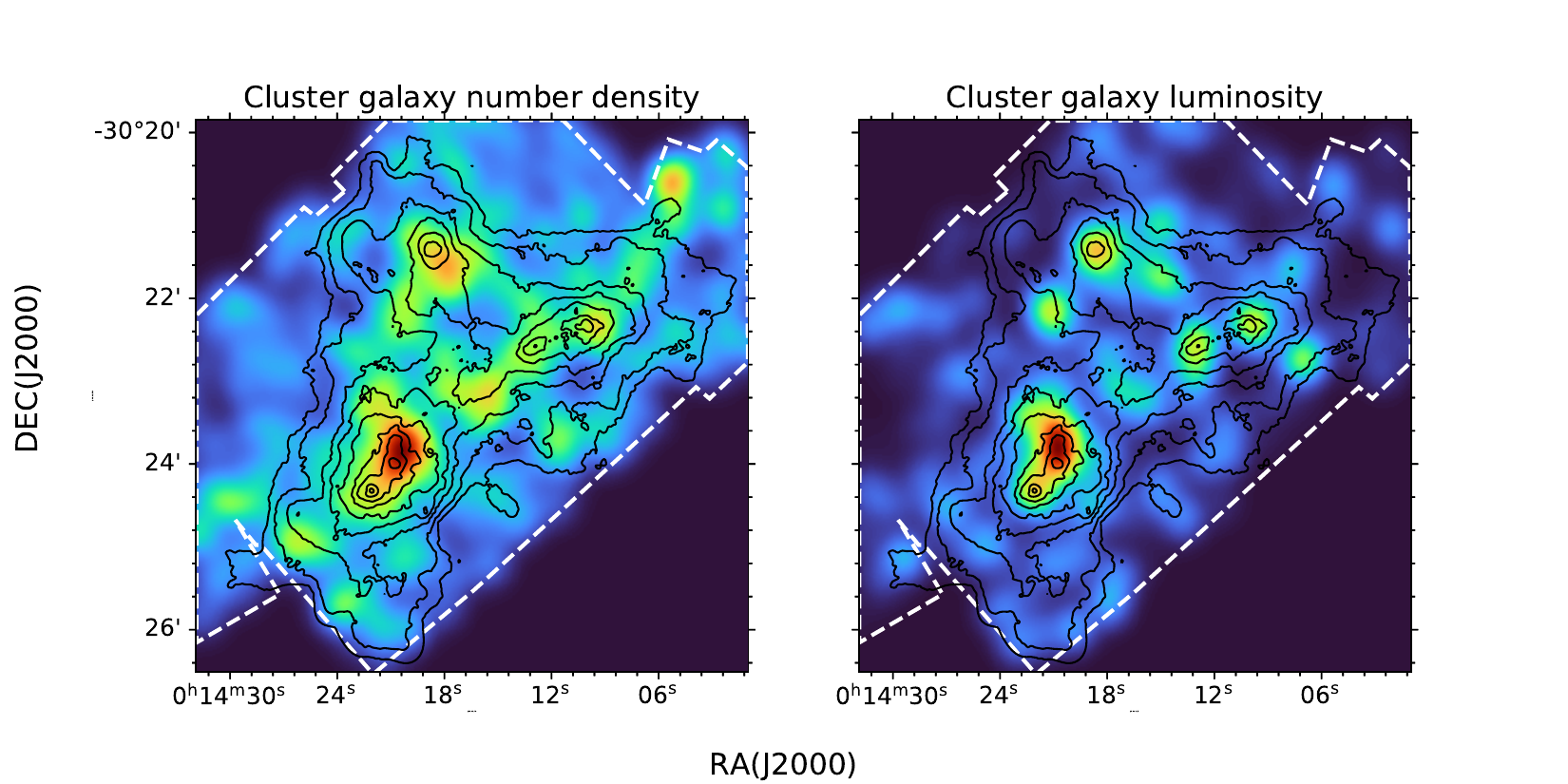} 
\caption{Mass contours overlaid on the cluster galaxy number and luminosity density maps. 
The left (right) panel shows the color map indicating the smoothed (Gaussian kernel with $\sigma\sim10\arcsec$) cluster galaxy number density (luminosity). The black solid lines represent the mass contours derived from our combined mass map, which are the same as shown in Figure~\ref{all_massmaps}. The white dashed line presents the footprint of the JWST observations.}
\label{density_map}
\end{figure*}

We present our mass reconstruction results in Figure~\ref{all_massmaps}. 
The overall mass structure of A2744 within the current mass reconstruction field
revealed by the WL+SL datasets
is isosceles triangular and characterized by the three main subclusters: the northern (G3), northwestern (G1+G2), and southern (BCG-N + BCG-S) mass substructures.
The legs of the isosceles triangle (the distances between BCG-S and G2 and between BCG-S and G3) are $\mytilde1$~Mpc whereas the base (the distance between G3 and G2) is $\mytilde0.6$~Mpc.
The northwestern (southern) substructure is further resolved into two smaller peaks: G1 and G2 (BCG-N and BCG-S).
It is remarkable that the five strongest mass peaks are precisely aligned with the five most luminous cluster galaxies ($\lesssim 2\arcsec$), although {\tt MARS} is entirely blind to the cluster galaxy distribution and never uses the LTM assumption.

A comparison of the WL+SL result with the WL-only (upper left) and the SL-only (upper right) ones delivers a few important takeaway messages.
First, the centroids of the five strongest mass peaks are well-constrained by either dataset. Although it is not surprising to see these alignments with the SL-only result, it is unprecedented that the WL data alone can constrain the mass centroids at this high-significance level
(S/N $\gtrsim 9\sigma$).
We believe that this is enabled by the unprecedentedly high WL source density ($\mytilde350$~arcmin$^{-2}$).
Second, the SL-only mass reconstruction provides strong constraints only within the SL regimes, which are confined to the $r\lesssim 0.3$~Mpc region in the southern subcluster,  the $r\lesssim 0.2$~Mpc region in the northwestern subcluster, and the $r\lesssim0.1$~Mpc region in the northern subcluster. Outside the SL regime, the mass density in the SL-only result is determined by the initial prior
(i.e., a flat convergence with $\kappa=0.1$ ).
We need the WL dataset to put meaningful constraints outside the SL regimes. 
Third, the WL dataset does not detect any significant mass peaks other than the aforementioned five mass peaks (BCG-N, BCG-S, G1, G2, and G3). \citet{2016MNRAS.463.3876J} presented WL+SL analysis and reported identifications of eight significant substructures in A2744. Among them, four (N, NW, S3, and core in their notation) coincide with our mass peaks. The other four are not
present in our mass reconstruction. Among these four, one substructure (Wbis in their notation) is located near the field boundary of the JWST footprint, and thus our result cannot be used to rule out its presence. 

We compare our mass map to the cluster member galaxy distributions in Figure~\ref{density_map}. 
We select cluster member candidates whose photometric redshifts are within the $0.28<z_{phot}<0.32$ range. We only select the objects that are brighter than 24 mag. We applied $3\sigma$ clipping and performed a linear fit in the color-magnitude diagram. The final member selection is made by identifying galaxies within $1\sigma$ of the best-fit relation. The F277W-F444W color is used because the combination clearly highlights the red sequence galaxies.
Since we already demonstrated that the five strongest mass peaks precisely coincide with the five brightest galaxies, good degrees of mass-light agreements are somewhat expected in this comparison. However, we note that the number density peaks when smoothed do not always fall exactly on the mass peaks. Although the cluster member catalog is incomplete, we believe that the offsets are primarily due to asymmetric galaxy distributions around the deepest potential wells. We suspect that the ongoing mergers may contribute to the asymmetry. We also observe that there are some luminosity/number density clumps, which do not have distinct mass counterparts. Perhaps, they are groups with low mass-to-light ratios or concentrations of galaxies that are not gravitationally bound and are only projected along the line-of-sight direction.

\subsection{Cumulative Projected Mass}\label{mass_measurement}
\begin{figure*}
\centering
\includegraphics[width=0.85\textwidth]{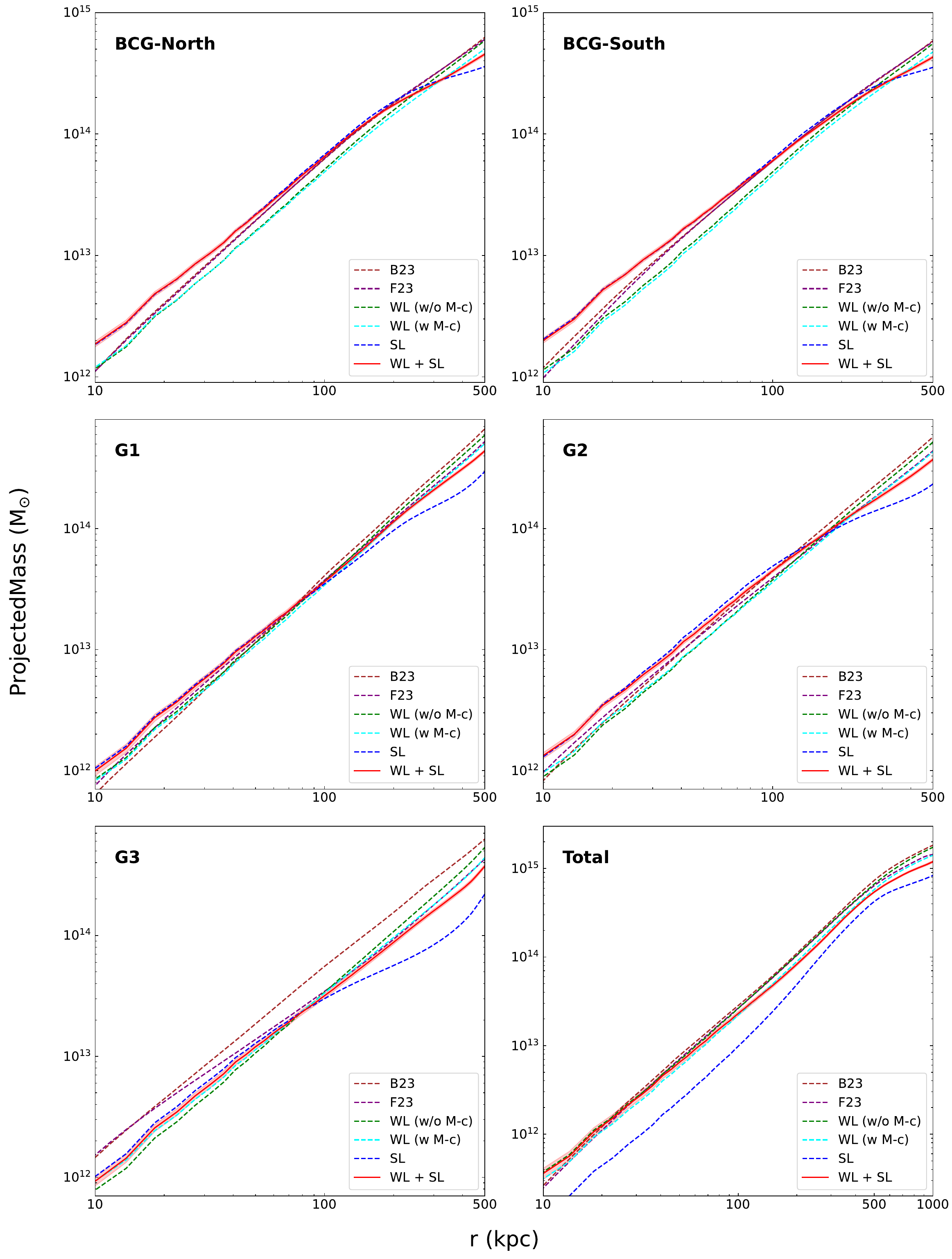} 
\caption{Cumulative projected mass profiles of A2744. We present the radial cumulative mass profiles of 5 halos (BCG-North, BCG-South, G1, G2, and G3). Additionally, we include the cumulative profile from the center of the field of view. The green (cyan) dashed lines indicate the cumulative mass profiles from the WL NFW profile fitting result without (with) the $M-c$ relation. The blue dashed lines represent the SL-only mass profiles derived from the current study. The purple and brown dashed lines represent the mass profiles from \citet{2023MNRAS.523.4568F} and \citet{2023ApJ...952...84B}, respectively. 
The red solid line and shaded regions display the profiles and the 1-sigma uncertainties of our combined (WL+SL) mass model, respectively.}
\label{all_radial_profile}
\end{figure*}

We present the cumulative mass profiles of the five mass peaks in Figure~\ref{all_radial_profile}. Also displayed is the total mass profile from the field center (RA=3.568514, DEC=-30.386321) roughly corresponding to the geometric center of the isosceles triangle defined by the three mass peaks: G3, G2, and BCG-S. 
The WL+SL and SL mass profiles are measured directly from the $\kappa$ map while the WL results are derived by simultaneously fitting five NFW profiles to the WL data with and without the mass-concentration ($M-c$) relation of \citet{2008MNRAS.390L..64D}; we cannot directly use the convergence map obtained from the WL-only result to estimate the mass because the $\kappa$ value in the SL regime is significantly underestimated, which is the combined effect of the mass-sheet degeneracy and regularization.
We refer readers to Appendix~\ref{NFW_fitting} for more details of the NFW fit.

Overall, the best-fit NFW profiles from WL yield smaller masses near the mass peak centers and larger masses at large radii ($r\gtrsim300$~kpc) than the main WL+SL results obtained directly from the convergence map. 
This implies that the densities at the mass peaks are significantly higher than the best-fit NFW predictions derived from our WL data. We provide two-dimensional comparisons on the issue in Appendix \ref{NFW_fitting}. The SL-only results are similar to the WL+SL ones in the SL regime ($r\lesssim100$~kpc) but are systematically lower at larger radii, where the lack of constraints makes the density default to the initial prior.

The WL+SL result shows that the total projected mass within $r=200$ kpc from BCG-North is $\sim1.73\times10^{14}M_{\odot}$, which is consistent with the values in \citet{2023ApJ...952...84B} and \citet{2023MNRAS.523.4568F}. However, for the other mass peaks, our mass model provides lower values than the parametric models. The projected mass within 200 kpc from G1, G2, and G3, are $\sim1.14\times10^{14}M_{\odot}$, $\sim1.15\times10^{14}M_{\odot}$, and $\sim8.77\times10^{13}M_{\odot}$, respectively. The projected mass of G3 has the largest difference from the value of \citet{2023ApJ...952...84B}. This is perhaps because there is only one multiple-image system around G3 and thus our free-form model cannot produce a sharp peak there.
Except for G3, the mass profiles from \citet{2023MNRAS.523.4568F}
are similar to the profiles derived from our WL-only best-fit NFW model, which are systematically lower than the ones from our main (WL+SL) model.
Our WL+SL model estimates the total projected mass within the $r=1$~Mpc aperture from the field center to be $\sim1.19\times10^{15}M_{\odot}$.

\subsection{Magnification}\label{magnification}
\begin{figure*}
\centering
\includegraphics[width=\textwidth]{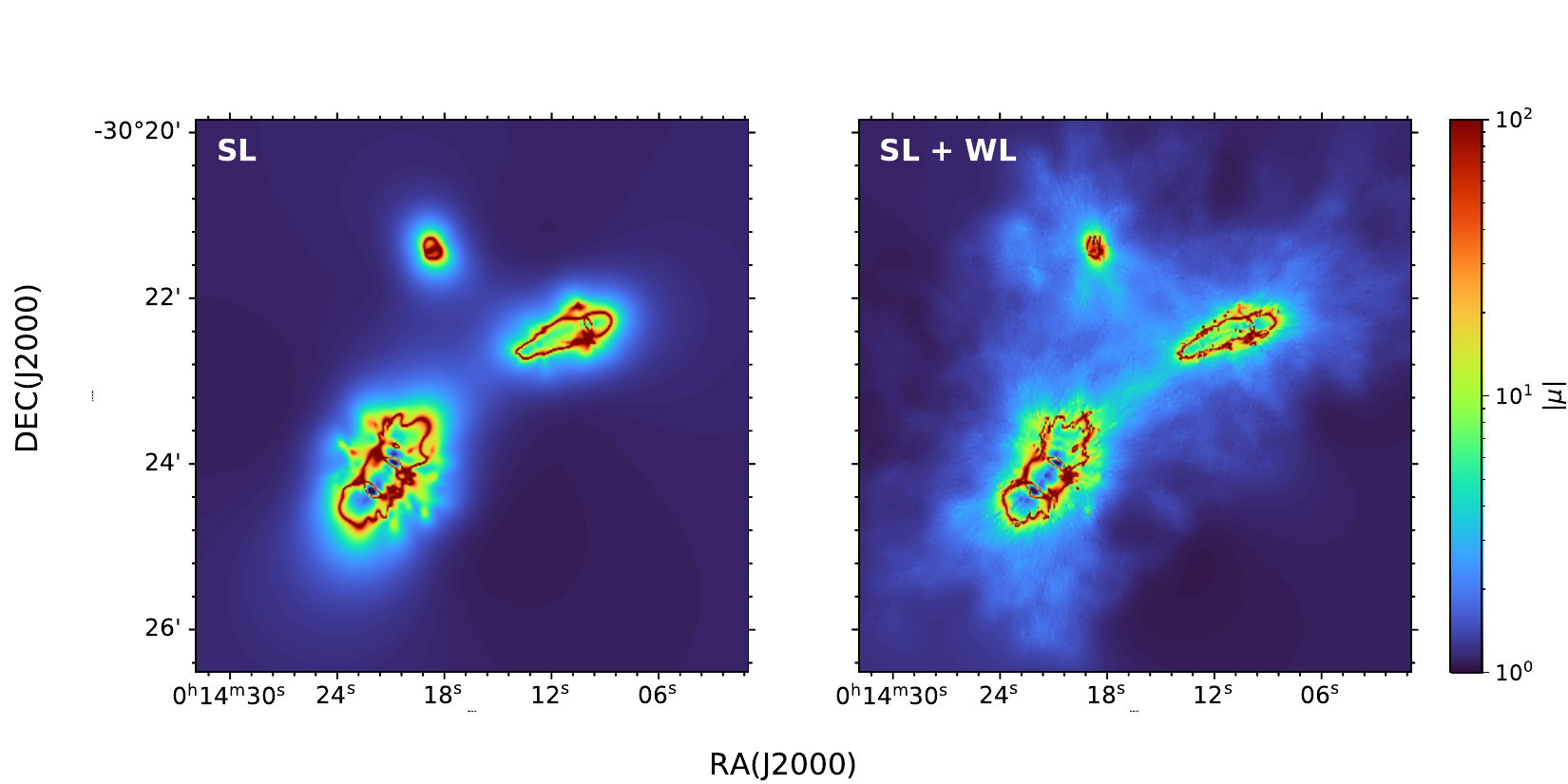} 
\caption{Magnification maps of the reconstructed lens models at the reference redshift $z_{s}=10$. The left (right) panel shows the magnification map from the SL-only (SL + WL) mass model. Unlike Figure~\ref{all_massmaps}, we do not apply the smoothing to the magnification map from the combined mass model.}
\label{magnification_map}
\end{figure*}

\begin{figure*}
\centering
\includegraphics[width=\textwidth]{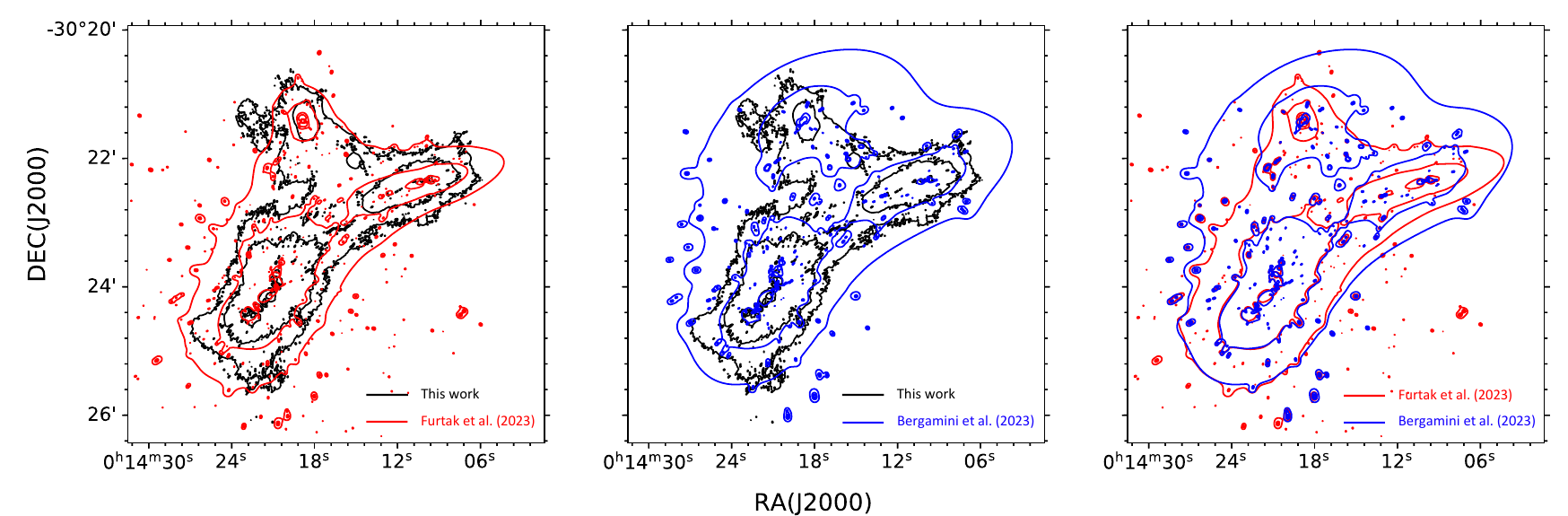}
\caption{Magnification comparison with the literature. The black contours show the magnifications from our combined mass map. The red and blue contours represent the magnifications from \citet{2023MNRAS.523.4568F} and \citet{2023ApJ...952...84B}, respectively. For each magnification, the inner (outer) contour indicates $|\mu| = 4 ~ (|\mu| = 2)$. The reference redshift is $z_{s}=10$.}
\label{compare_mag}
\end{figure*}

Figure~\ref{magnification_map} displays the magnification maps from the SL-only and SL + WL models. Similar to the mass map, the magnification map is
characterized by the three main critical curve loops. The overall structures of the critical curves are in broad agreement with those reported in the literature \citep[e.g.,][]{2023MNRAS.523.4568F, 2023ApJ...952...84B}. However, the resolution limitation and lack of images around compact halos cause {\tt MARS} to exhibit some lack of detail near the cluster member galaxies.  

The shapes of the critical curves in the SL-only model are similar to those in the WL+SL model. These similarities are expected because the critical curves are primarily constrained by the SL dataset. However, the WL+SL result provides significantly higher and more detailed magnification in the outskirts. This is because the lack of constraints makes the SL-only model predict much lower and simpler mass densities (defaulting to the initial prior) outside the SL regime.

In Figure~\ref{compare_mag}, we also compare our magnification map with the results from \citet{2023MNRAS.523.4568F}\footnote{\url{https://jwst-uncover.github.io/DR1.html}} and \citet{2023ApJ...952...84B}\footnote{\url{https://www.fe.infn.it/astro/lensing/}}, who kindly made their results publicly available. Although the overall morphology is similar, the details are significantly different among three magnification maps. The magnification contours from \citet{2023MNRAS.523.4568F} and \citet{2023ApJ...952...84B} extend wider, predicting higher magnification values in the outskirts. 
Especially, \citet{2023ApJ...952...84B} suggests broad higher magnification distributions around G3. We attribute this difference to our mass profile constrained by the WL data decreasing faster than the parametric descriptions used in \citet{2023MNRAS.523.4568F} and \citet{2023ApJ...952...84B}.

\section{Discussion} \label{sec:discuss}

\subsection{Robustness Test of the Mass Model}\label{robust_lens_model}
A robust mass model is expected to accurately reproduce observed SL and WL features without overfitting. Here we assess the quality of our mass model using the following four metrics: lens-plane scatters, lens-plane image reconstructions, per galaxy shear predictability, and tangential shear profiles. It is important to note that satisfactory performance assessed by these metrics is only a necessary condition, not a sufficient condition, for a robust mass model.

\subsubsection{Lens Plane Scatters of Multiple Images}\label{lens_plane_rms}
\begin{figure}
    \includegraphics[width=0.46\textwidth]{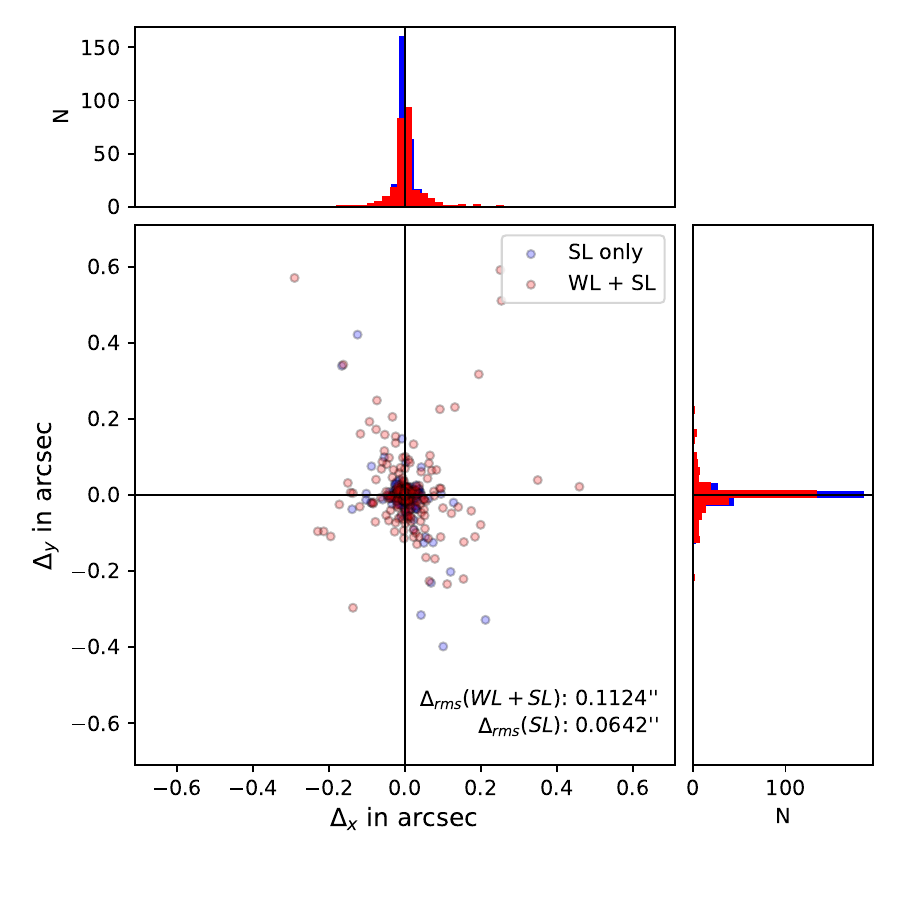} 
    \caption{Lens plane scatters between the observed and predicted locations of multiple images. $\Delta_{x(y)}$ represents the deviation from the observation along the x-axis (y-axis). The red (blue) dots indicate the lens plane scatter distributions derived from the WL + SL (SL-only) mass map. $\Delta_{rms}$ is the rms value of the total lens plane scatters  (see equation~\ref{eqn_rms}).} 
    \label{fig:img_scatter_comparison}
\end{figure}

One of the most common metrics used for the evaluation of SL mass models is the difference between the observed and predicted locations of multiple images on the image plane. We compute the rms of the position differences between the observed and predicted multiple images from our mass model using the following: 
\begin{equation}\label{eqn_rms}
    \Delta_{rms}=\sqrt{\frac{1}{M}\sum_{m=1}^{M}|\bm{\theta}_{truth,m}-\bm{\theta}_{model,m}|^{2}},
\end{equation}
where $M$ represents the total number of multiple images, and $\bm{\theta}_{truth,m}$ and $\bm{\theta}_{model,m}$ are the locations of the observed and predicted multiple images for the $m^{th}$ image, respectively.

In Figure~\ref{fig:img_scatter_comparison}, we plot the distributions of lens-plane scatter. The rms value for the SL-only (SL + WL) mass model is $\Delta_{rms}=0\farcs06 ~(0\farcs11)$. 
The SL-only mass model in this study yields an rms value slightly higher than our previous result \citep[][$\Delta_{rms}=0\farcs05$]{2023ApJ...951..140C}, where the mass reconstruction was limited to the main cluster region (within the single ACS pointing).
The scatter increases approximately by a factor of two when the WL data are included (from $0\farcs05$ to $0\farcs11$). 
This increase is not surprising because the inclusion of the $\chi^2_{WL}$ term in equation~\ref{total_eqn} effectively lowers the weight on the $\chi^2_{SL}$ term under the same regularization.
Nevertheless, we emphasize that this rms value $(0\farcs11)$
is still a factor of four lower than those in other JWST-based SL-only studies; for instance, the scatters are $\Delta_{rms}=0\farcs51$ and $0\farcs43$ for \citet{2023MNRAS.523.4568F} and \citet{2023ApJ...952...84B}, respectively.
Since the JWST image informs us of the location of the multiple images within the accuracy of a few pixels, our relatively small scatter ($0\farcs112$) should not be attributed to overfitting.

\subsubsection{Lens-Plane Image Reconstruction}\label{lens_plane_image_recon}
\begin{figure*}
\centering
\includegraphics[width=0.9\textwidth]{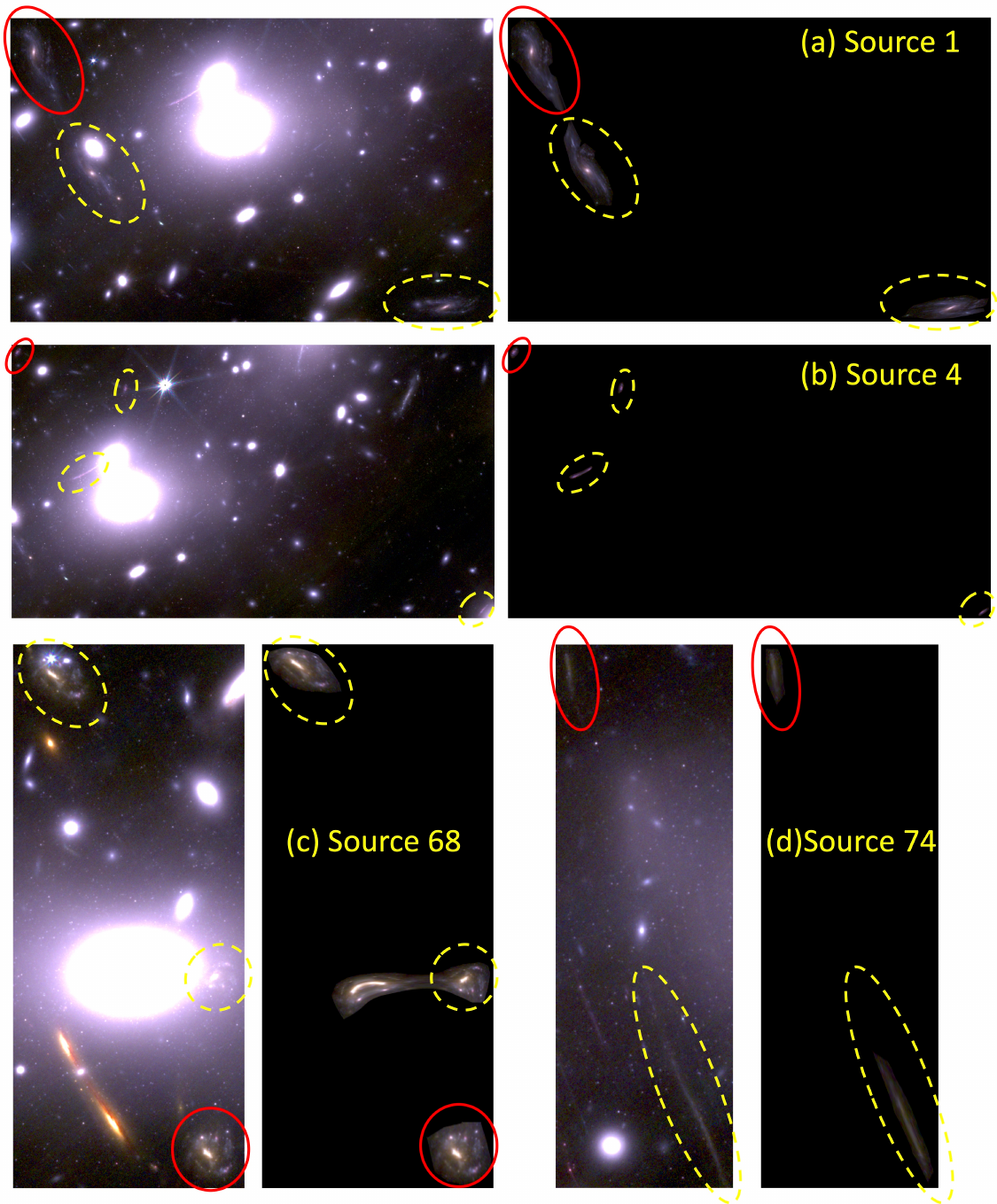} 
\caption{Lens-plane image reconstructions. The left panels show the observed multiple images, with red circles indicating the selected images for reconstruction. The yellow dashed circles represent the locations of multiple images. Right panels present the reconstructed images in the lens plane, with dimensions matching those of the left panels. The color-composite images are the same as shown in Figure~\ref{newly_identified}.}
\label{mulitple_recon}
\end{figure*}

Although the lens-plane scatter metric (\textsection\ref{lens_plane_rms}) provides a useful statistic to assess the quality of the lens model, it does not inform us of the robustness of the lens model on small scales in the neighborhood of the multiple image positions. In particular, overfit models from free-form approaches fail to recover lens-plane galaxy morphologies reliably because of high-frequency noise.

Figure~\ref{mulitple_recon} displays the reconstructed multiple images in the lens plane from our WL+SL mass model. We choose the four systems that exhibit highly distorted multiple images because they are more sensitive to details in the mass distribution.
In general, it is easier to reconstruct the morphology of the system with more resolved features since the mass reconstruction utilized them.
Sources 1 (a) and 68 (c) are the systems where we identified 4 knots in the mass reconstruction while sources 4 (b) and 74 (d) are the ones for which we only used their centroids.
Although the images with multiple knots provide better reconstructions, the case with a single constraint also yields good results. We performed these tests with other multiple image systems and verified that their reconstruction qualities are similar.

Therefore, we conclude that our lens model is stable on small scales in the vicinity of the multiple image positions even after we require the mass model to account for all WL features.

\subsubsection{Shear Predictability per Galaxy}\label{chi2_WL_galaxy}
\begin{figure}
    \includegraphics[width=0.46\textwidth]{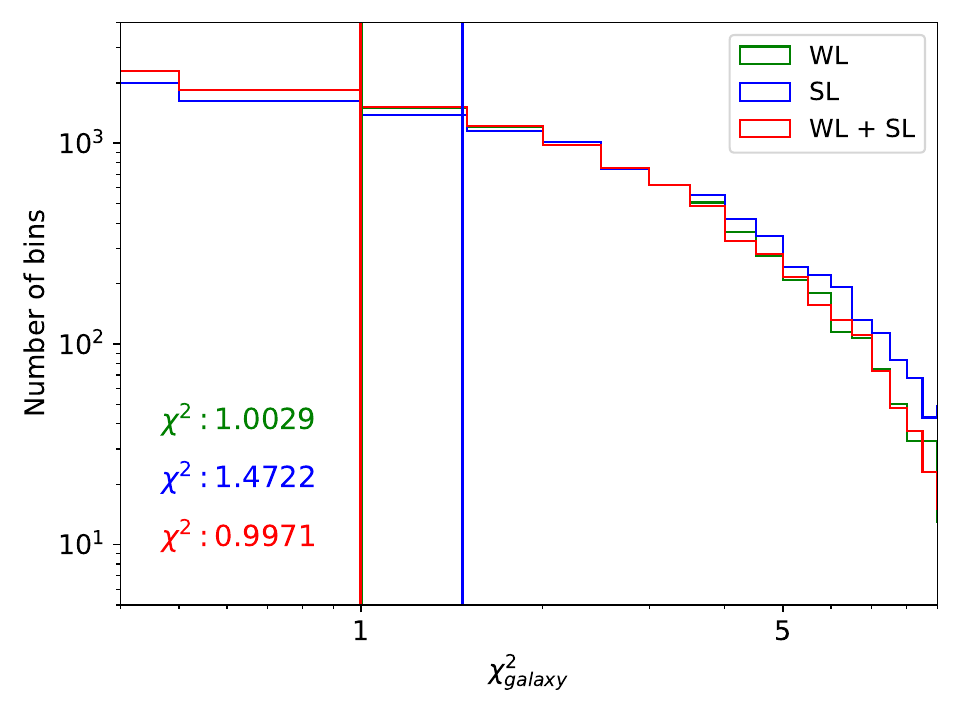} 
    \caption{Distributions of per-galaxy $\chi^2_{WL}$. The green (blue) histograms indicate the distributions of the per-galaxy $\chi^2_{WL}$ obtained from the WL-only (SL-only) mass map. The red histograms represent the distributions of the per-galaxy $\chi^2_{WL}$ from the WL+SL mass map. The vertical lines correspond to the mean per-galaxy $\chi^2_{WL}$ values from each mass map, indicated by the numbers at the lower left (matching the color of each histogram).}
\label{chi2_WL_distribution}
\end{figure}

Having verified that our combined mass model reproduces the SL features in terms of the multiple image positions and morphologies in \textsection\ref{lens_plane_rms} and \textsection\ref{lens_plane_image_recon}, here we discuss how much the predicted shears at the galaxy positions are consistent with the observations.
Since each galaxy's ellipticity measurement contains its intrinsic shape and measurement noise, as well as the shear, it is important to include them in our judgment of the goodness of the fit. Since Equation~\ref{chi_square_WL} is already designed to accommodate such a need, we decide to utilize it and adopt the distribution of the normalized squared residual per galaxy ellipticity component (hereafter we refer to it as per-galaxy $\chi^2_{WL}$) as our metric.

Figure~\ref{chi2_WL_distribution} displays the distributions of the per-galaxy $\chi^2_{WL}$ measured for the WL-only, SL-only, and WL+SL mass reconstructions. 
Both WL-only and WL+SL models provide a mean per-galaxy $\chi^2_{WL}$ close to unity whereas the value is $\mytilde50$\% higher for the SL-only model.
Since the SL-only models from \citet{2023MNRAS.523.4568F} and \citet{2023ApJ...952...84B}, which use the JWST observations, are publicly available, we retrieved the models and computed their per-galaxy $\chi^2_{WL}$ with our WL data. We found that the mean per-galaxy $\chi^2_{WL}$ from \citet{2023MNRAS.523.4568F} and \citet{2023ApJ...952...84B} are $\mytilde40$\% and $\mytilde32$\% higher, respectively, than our WL+SL or WL-only result.
This illustrates that our final WL+SL model robustly reproduces the WL features as well as the SL features. Also, this serves as an important lesson that a complete mass model within the current JWST A2744 field requires WL constraints.

\subsubsection{Reduced Tangential Shear Profile}\label{radial_tangential}
\begin{figure}
    \includegraphics[width=0.47\textwidth]{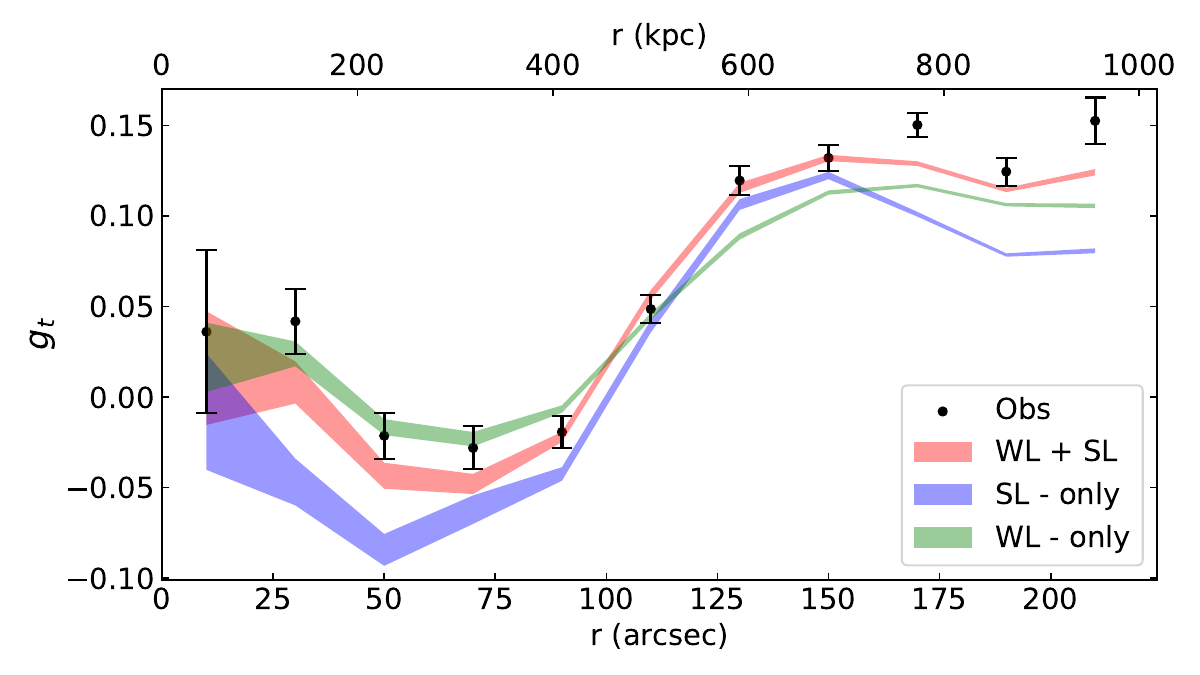} 
    \caption{Radial reduced tangential shears comparison. We compute the radial tangential shears from the center of the field of view. The dots indicate the radial reduced tangential shears. The errorbars and shaded regions show the standard uncertainties. The black, red, blue, and green samples are obtained from the observations, combined model, SL-only model, and WL-only model, respectively.}
\label{rad_tan_compare}
\end{figure}

The reduced tangential shear is a measure of how much the shapes of background galaxies are tangentially aligned with respect to a reference point. Here we adopt the center of the mosaic image as our reference because it maximizes the radius where the measurement is derived from a complete circle. The reduced tangential shear $g_t$ is evaluated via the following equation:
\begin{equation}\label{eqn_tan_shear}
    g_{t}=-g_{1} \cos 2\phi - g_{2} \sin2\phi,
\end{equation}
where $g_{1(2)}$ denotes the first (second) component of the reduced shear, and $\phi$ represents the position angle of the object measured counterclockwise from the reference axis. The amplitude of the reduced tangential shear is given by [$\bar{\kappa}(<r)-\kappa(r)]/[1-\kappa(r)]$ and is sensitive to the overall shape of the radial mass profile.
Therefore, it is possible that a mass model that performs well in the above per-galaxy $\chi^2_{WL}$ test performs poorly in this tangential shear test, and vice versa.

Figure~\ref{rad_tan_compare} displays the comparison of observed tangential shears with the model prediction. We remind the reader that the radial behavior of the current tangential shear is different from those of the typical cases in the literature because of two reasons. First, the reference point is at the center of the mass reconstruction field, which is near the geometric center of the three substructures of A2744. Second, the mass distribution of A2744 is by and large trimodal.
The relatively low projected mass density near the field center makes the amplitude of the tangential shears remain low until the radius reaches $r\sim600$~kpc, which is approximately the mean distance from the field center to the three substructures. 

Some noticeable deviations from the observation are present in the SL-only and WL-only mass models.
The SL-only mass model predicts that the reduced tangential shears are initially more negative at $r\lesssim400$~kpc, reach levels similar to the observation at $500~\mbox{kpc}\lesssim r \lesssim700~\mbox{kpc}$, and become significantly lower at larger radii ($r\gtrsim800$~kpc). This pattern is attributed to the absence of the multiple image constraints at $r\lesssim400$~kpc or $r\gtrsim800$~kpc, where our SL-only mass reconstruction tends to default to the initial prior.
The WL-only mass model predicts the tangential shears well up to $r\lesssim500$~kpc. However, beyond this, its prediction is systematically lower than the observation. This is due to the fact that our maximum entropy regularization makes it difficult for the WL-only mass reconstruction to reach high $\kappa$ values in the SL regime since sharper mass peaks lower entropy. 
The WL+SL mass model provides the best predictions, which match the observed values across the entire range. 

The current tangential shear test illustrates that it is important to incorporate both SL and WL datasets when a high-fidelity wide-field mass reconstruction is desired in a massive cluster. As observed, WL-only mass reconstruction is biased because the WL data alone cannot adequately inform us of the mass distribution around the extremely high-density regions. On the other hand, SL-only reconstruction is only robust in the SL regime, where multiple images are densely distributed. Although this issue can be somewhat mitigated by assuming that the mass profile follows some analytic descriptions, in post-merger clusters, such as A2744, the assumption may diminish our ability to learn how the mass profiles are affected by the sub-halo collisions. 

\subsection{Comparison with Previous Studies and Merging Scenarios}\label{merging_scenario}

\begin{figure*}
\centering
\includegraphics[width=\textwidth]{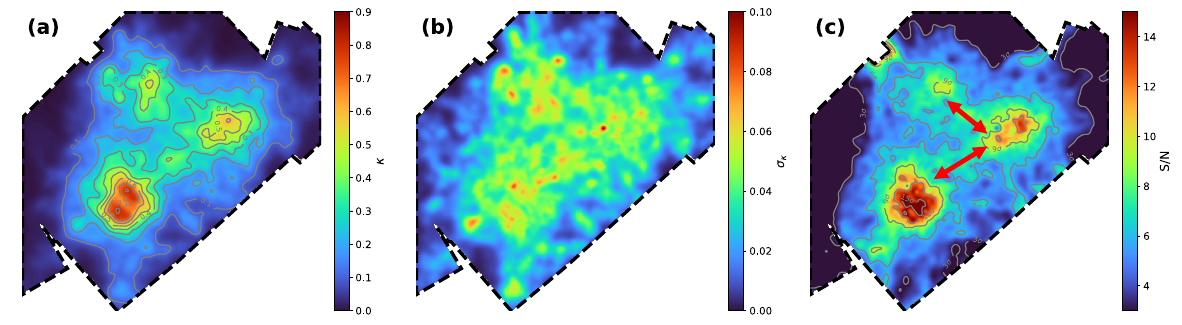} 
\caption{Mass, uncertainty, and signal-to-noise maps of the WL-only mass model. The dashed black lines indicate the footprint of the JWST observations. All three maps are smoothed with a Gaussian kernel of $\sigma=2\arcsec$.
(a) WL-only mass map. The contour labels show the convergence $\kappa$.
(b) Uncertainty of the WL-only mass model derived from 1000 bootstrap realizations. 
(c) S/N map of the WL-only mass model. The contour levels are $3\sigma,~6\sigma,~9\sigma,~12\sigma$, and~$15\sigma$. The red arrows indicate the mass bridges.
}
\label{WL_SL_SN}
\end{figure*}

\begin{figure*}
\centering
\includegraphics[width=\textwidth]{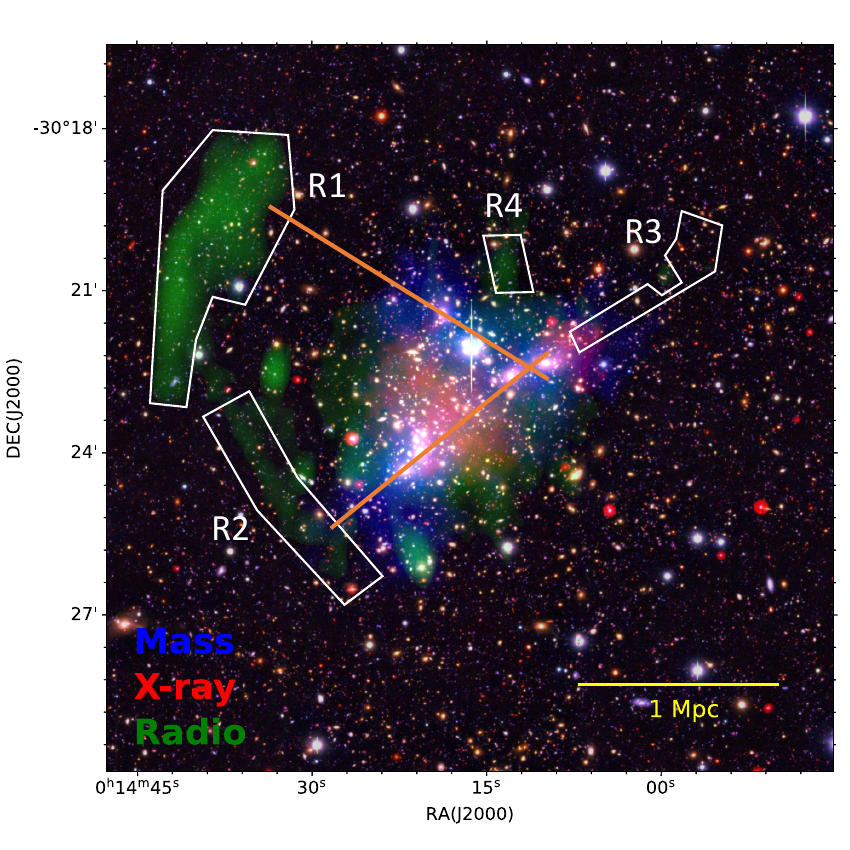} 
\caption{WL + SL Mass reconstruction of A2744. The blue intensity region represents the mass distributions from the WL + SL mass map. The red intensity region corresponds to X-ray surface brightness (OBSID: 7915, 8477 and 8557; PI: J. Kempner). The green region displays radio continuum from radio observations \citep[][; GMRT data at 325MHz]{2013A&A...551A..24V, 2019MNRAS.489..446P}. The color composite image is created using the Subaru/Suprime-Cam observations, with the z band for red, R band for green, and B band for blue \citep[][]{2022AAS...24021405F}. The areas enclosed with white solid lines indicate the four known radio relics \citep{2017ApJ...845...81P, 2021A&A...654A..41R}. The orange solid lines show the expected merger axes from the mass bridges. The field of view is $13\farcm46\times13\farcm46$.}
\label{WL_SL_color}
\end{figure*}

Due to its rich and puzzling substructures, A2744 was introduced with its nickname ``Pandora's cluster" \citep{2011MNRAS.417..333M}.
One of the notable features highlighted in \cite{2011MNRAS.417..333M} was the ``ghost" clump with no apparent correlation with the cluster galaxies. In addition, the authors reported large offsets between BCGs and mass peaks, which were supported by \citet{2016ApJ...817...24M}.
However, other studies \citep[e.g.,][]{2016MNRAS.463.3876J, 2023MNRAS.523.4568F, 2023ApJ...952...84B} based on SL-only or WL + SL datasets found neither significant offsets between BCGs and mass peaks nor the ghost clump. A caveat in the latter studies is that the mass models are reconstructed based on the LTM assumption, and thus the possibility of mass peaks with considerable departure from the galaxies is not extensively explored.
The current study is the first free-form mass reconstruction of A2744 based on WL+SL with no LTM assumption. Our JWST result supports neither the existence of the ghost clump nor the mass-galaxy offsets, although our analysis is completely blind to the locations of the cluster galaxies in A2744.

Intracluster stars and globular clusters have been suggested as visible tracers of dark matter \citep[e.g.,][]{2010ApJ...717..420J,2020MNRAS.494.1859A, 2022ApJS..261...28Y,2022ApJ...940L..51M,2023arXiv230103629D} if their formation occurs at high redshifts \citep{2018ApJ...862...95K,2023Natur.613...37J, 2023MNRAS.523...91W}. Recently, \citet{2023arXiv230714412H} identified more than 10,000 intracluster globular clusters in A2744 with the same JWST imaging data that we used in the current study. The spatial distribution of the intracluster globular clusters in A2744 closely follows the lensing-based mass map presented in the current and other studies \citep{2023MNRAS.523.4568F, 2023ApJ...952...84B}.

The scrutiny of our mass map hints at the existence of two ``mass bridges": one between NW and S, and the other between N and NW. These density enhancements are mainly constrained by the WL dataset, which provides unprecedentedly high source density ($\mytilde350$~arcmin$^{-2}$).
To estimate the significance of these mass bridges, we compute an uncertainty of the WL-only mass map (Figure~\ref{WL_SL_SN}). We reconstruct 1000 mass maps from bootstraps of the original WL catalog and measure the standard deviation (Figure~\ref{WL_SL_SN}b), which we adopt as the uncertainty map. The S/N map (Figure~\ref{WL_SL_SN}c) is obtained by dividing the WL-only mass map by this uncertainty map.
The significance of the contours outlining these mass bridges is  $\mytilde6.0~\sigma$ according to the S/N map. Since A2744 is one of the most massive clusters known to date, the mass density within the current JWST field should be significantly positive and thus the true background level can be estimated only from studies with much larger fields.
Nevertheless, as a conservative measure, we also evaluated the significance of the mass bridges with respect to
the background level estimated within the reconstruction field.
When adopting the outermost contour level of the mass map (Figure~\ref{WL_SL_SN}a) as the baseline, we find that the significance decreases to $\mytilde4.0~\sigma$, which implies that the mass bridge features are still high in this conservative measure.
The mass bridges may be interpreted as arising from the mergers since numerical simulations show that mass bridges develop between the two clusters after the core passage.

Interestingly, the orientations and locations of the two largest radio relics in A2744 are consistent with the hypothesized merger axes inferred by the two mass bridges (Figure~\ref{WL_SL_color}). The brighter relic (R1) is located $\mytilde1$~Mpc away from the northern clump and is on the hypothesized merger axis connecting NW and N. Also, the orientation of the relic is perpendicular to the merger axis.
The fainter relic (R2), $\mytilde0.5$~Mpc southeast of the southern (main) clump, is also on the axis connecting NW and S, and again its orientation is orthogonal to the axis.

However, despite the above intriguing possibility, complete reconstruction of the merging scenario of A2744 is still challenging. First of all, A2744 consists of
at least five massive substructures, which provide too many degrees of freedom in plausible merging scenario reconstruction. Furthermore, the X-ray morphology of A2744 is complex and contains many substructures significantly dissociated from the galaxy distributions. Although similar degrees of gas-galaxy offsets have been observed in other binary mergers \citep[e.g.,][]{2006ApJ...648L.109C, 2016A&A...594A.121P}, the A2744 complexity cannot be explained by such binary encounters.

\section{Conclusion} \label{sec:conclusion}
Leveraging 286 multiple images and a WL source density of $\sim350~\rm arcmin^{-2}$, we have presented a new WL + SL mass model of A2744 using the {\tt MARS} free-form algorithm. For the WL analysis, we carefully modeled the PSF and measured the ellipticities of the background sources, whose selection is based on photometric redshifts. For the SL constraints, we compiled multiple images from the literature and also identified new multiple-image candidates. 

Our WL+SL mass reconstruction provides the highest-resolution mass map of Abell 2744 within the $\mytilde1.8$~Mpc$\times1.8$~Mpc field to date, revealing the giant isosceles triangular structure characterized by two legs of $\mytilde1$~Mpc and a base of $\mytilde0.6$~Mpc. 
While our algorithm {\tt MARS} remains entirely unaware of the distribution of cluster galaxies, the resulting mass reconstruction remarkably traces the brightest cluster galaxies, with the five strongest mass peaks coinciding with the five most luminous cluster galaxies.
The most remarkable features of our mass reconstruction include the two mass bridges: one connecting N and NW and the other connecting NW and S. These 4~$\sigma$ features are interpreted as arising from the on-going mergers because the merger axes defined by the features are consistent with the positions and morphologies of the two brightest radio relics in A2744.

We support the robustness of our mass model with various tests involving lens-plane position scatters, lens-plane morphology reconstruction, per-galaxy $\chi^2_{WL}$ statistics, and tangential shear profiles. We demonstrate that the WL+SL mass model performs well against all these tests while the performance is not satisfactory when only one of the two (WL and SL) datasets is used for the model construction. The comparison of the current result with those in previous studies shows some important differences in mass profiles and magnifications in both SL and WL regimes. We attribute them to both the utilization of the WL data outside the SL regime and the free-form {\tt MARS} algorithm.

The present study demonstrates that with the advent of JWST era, the cluster mass reconstruction combining WL and SL has now emerged as a critical and also practical channel to robustly measure the mass distributions of the massive clusters from their cores to the outskirts. This will enhance our comprehension of cluster physics, dark matter properties, and reionization-epoch galaxies.

We express our gratitude to Kyle Finner for providing the excellent Subaru/Suprime-Cam color image and contributing to the discussion of the merging scenario of A2744. Our gratitude goes to Pietro Bergamini for sharing maps from the lens model. We also extend our thanks to the UNCOVER team for making their photometric redshift catalog and reduced JWST images publicly available.
This work is based on observations created with NASA/ESA/CSA JWST and downloaded from the Mikulski Archive for Space Telescope (MAST) at the Space Telescope Science Institute (STScI).
The GLASS-JWST NIRCam observations analyzed can be accessed via \dataset[doi:10.17909/mrt6-wm89]{https://doi.org/10.17909/mrt6-wm89}. The JWST UNCOVER NIRCam observations analyzed can be accessed via \dataset[doi:10.17909/zn4s-0243]{https://doi.org/10.17909/zn4s-0243}.
The JWST DDT program 2756 NIRCam observations analyzed can be accessed via \dataset[doi:10.17909/te6f-cg91]{https://doi.org/10.17909/te6f-cg91}.
M. J. Jee acknowledges support for the current research from the National Research Foundation (NRF) of Korea under the programs 2022R1A2C1003130 and RS-2023-00219959.

\software{Astropy \citep{astropy2013,2018AJ....156..123A,2022ApJ...935..167A}, Matplotlib \citep{matplotlib2007}, NumPy \citep{harris2020array}, PyMultiNest \citep{2014A&A...564A.125B}, PyTorch \citep{NEURIPS2019_bdbca288}, SciPy \citep{scipy2020}, SEXtractor \citep{1996A&AS..117..393B}, \texttt{WebbPSF} \citep{10.1117/12.925230, 10.1117/12.2056689}}

\appendix
\section{PSF Modeling}\label{dif_psf_results}
\begin{figure*}
\centering
\includegraphics[width=\textwidth]{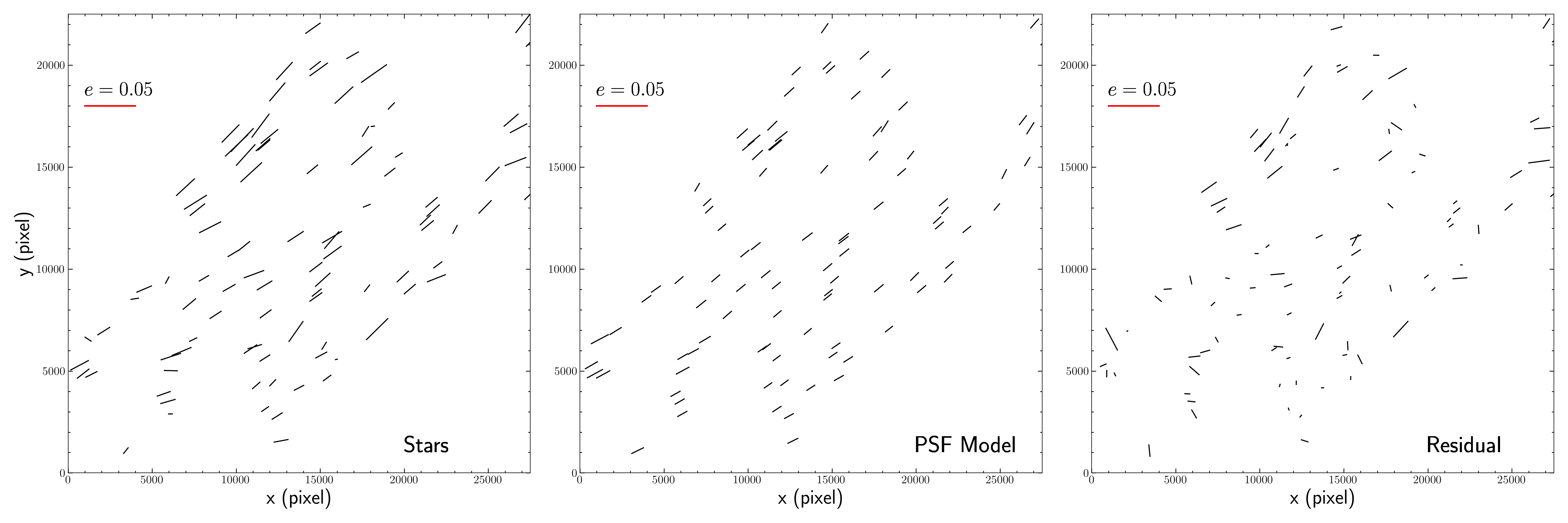}
\includegraphics[width=\textwidth]{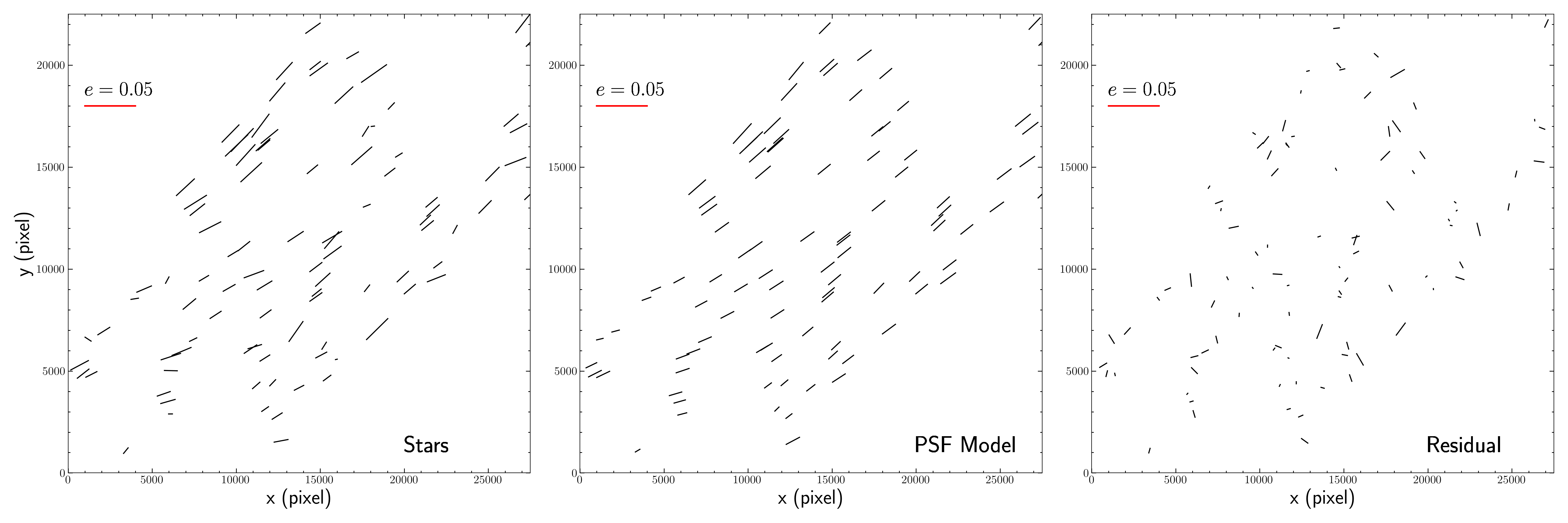}
\caption{Whisker plots showing ellipticity direction and magnitude at mosaic star positions for the two different PSF models. The upper (lower) three plots, from left to right, show star ellipticities, \texttt{WebbPSF} model (empirically obtained PSF model) ellipticities, and the residual ellipticities from subtracting the model ellipticities from the star ellipticities.}
\label{whiskers}
\end{figure*}

Two different PSF modeling methods were tested when analyzing the A2744 data. The first is described in Section~\ref{psf_modeling}, while the other process closely follows that from \citet{2023ApJ...953..102F}. This second, empirically obtained PSF model more accurately reproduces the observed PSF ellipticity at the locations of stars in the field, as seen by the smaller magnitude and more randomly oriented ellipticity residuals in Figure~\ref{whiskers}. However, this modeling technique makes the assumption that the mosaic PSF is continuous despite being the result of combining numerous exposures from various observation dates. Additionally, the empirically obtained PSF model is subject to the possibility of overfitting or underfitting the data based on the choice of polynomial order.

\begin{figure*}
\centering
\includegraphics[width=\textwidth]{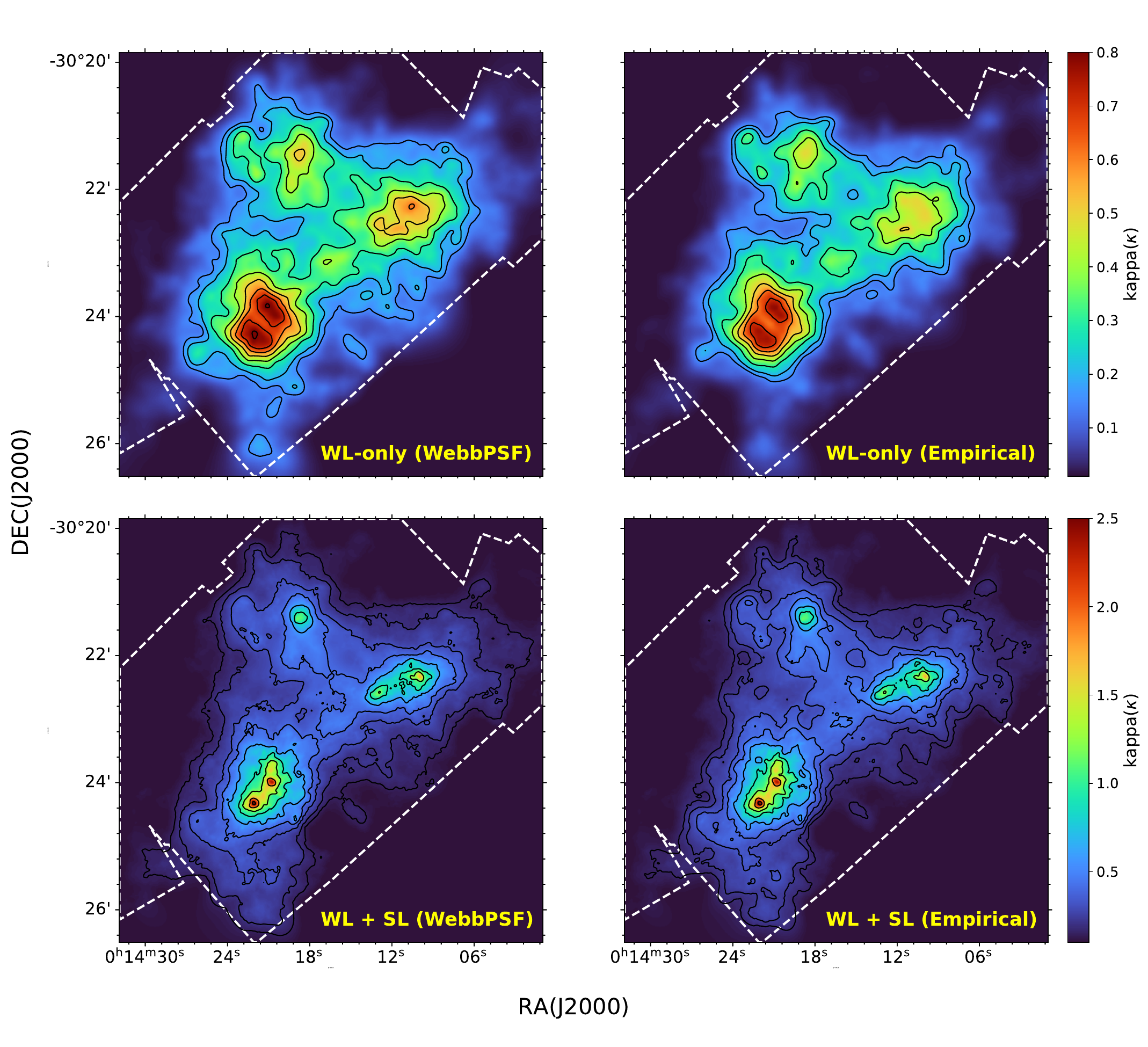} 
\caption{Mass contours of A2744 overlaid on the mass maps from the different PSF models. The upper (lower) row shows the WL-only (WL+SL) mass maps. The black solid lines represent the mass contours derived from our result mass maps, which are the same as shown in Figure~\ref{all_massmaps}. The white dashed line presents the footprint of the JWST observations.}
\label{mass_comparison_dif_psf}
\end{figure*}

\begin{figure}
    \centering
    \includegraphics[width=0.5\textwidth]{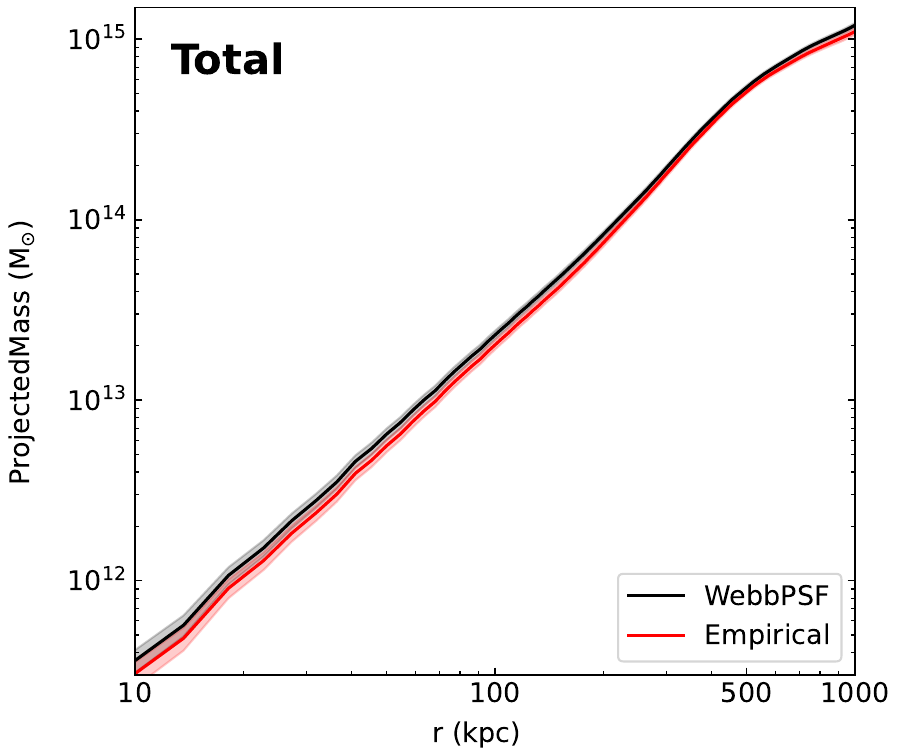}
    \caption{Cumulative projected mass profile of A2744 from the center of the field of view (Same as in Figure~\ref{all_radial_profile}). The black (red) solid line and shaded regions indicate the cumulative mass profile and the 1-sigma uncertainties of the combined mass model using \texttt{WebbPSF} (empirically obtained PSF).}
\label{radial_mass_comp_dif_psf}
\end{figure}

The \texttt{WebbPSF} model generally produces larger magnitude residual ellipticities at mosaic star positions, which in some cases maintain the observed alignment. However, given that the A2744 data was collected on different occasions, modeling the PSF for each input exposure may be more robust in terms of reproducing PSF characteristics away from observed star locations. Additionally, we find that there are no significant differences in the resulting mass maps utilizing the two different PSF models for WL catalogs. Figure~\ref{mass_comparison_dif_psf} presents WL-only and WL+SL mass maps using WL catalogs derived from the two PSF models. While there are some small differences in details, the overall mass distributions are highly consistent. The presence of mass bridges is also identified when we use the WL catalog from the empirically obtained PSF model. Furthermore, elongations of mass distributions remain regardless of which PSF model is used. 

Similar to the mass distributions, the total mass measured when using the different models remains consistent. We plot cumulative projected mass profiles of WL+SL mass maps from the center of the field of view in Figure~\ref{radial_mass_comp_dif_psf}. The cumulative masses derived from the two PSF models show agreement within 1-sigma uncertainties. The total projected mass within the $r=1$ Mpc aperture from the field center to be $\sim1.19\times10^{15}M_{\odot}$ and $\sim1.10\times10^{15}M_{\odot}$ from \texttt{WebbPSF} and the empirically obtained PSF model, respectively. Given the assumption of continuity made with the empirically obtained PSF model, the ambiguous choice of polynomial order, and the similarity in the final WL results between the two models, the \texttt{WebbPSF} model was chosen for the final WL analysis.

\section{Substructure Properties}\label{NFW_fitting}
\begin{deluxetable*}{cccccccc}\label{NFW_fitting_result_Mc} 
\tablecaption{NFW profile fitting results with $M-c$ relation}
\tablehead {
\colhead{} &
\colhead{} &
\colhead{WL} &
\colhead{} &
\colhead{} & 
\colhead{} &
\colhead{WL + SL} &
\colhead{}
\\ \cline{2-4}  \cline{6-8} 
\colhead{Component} &
\colhead{$M_{200}$} &
\colhead{$c_{200}$} &
\colhead{$R_{200}$} &
\colhead{} & 
\colhead{$M_{200}$} &
\colhead{$c_{200}$} &
\colhead{$R_{200}$}
\\
\colhead{} &
\colhead{($10^{14}M_{\odot}$)} &
\colhead{} &
\colhead{(kpc)} &
\colhead{} &
\colhead{($10^{14}M_{\odot}$)} &
\colhead{} &
\colhead{(kpc)}
}
\startdata
BCG-North &  $4.16\pm0.30$ & $3.313^{+0.018}_{-0.021}$ & $1385^{+35}_{-30}$ &  & $2.084\pm0.032$ & $3.5104\pm0.0045$ & $1100.6\pm5.6$ \\
BCG-South &  $2.12^{+0.20}_{-0.26}$ & $3.508\pm0.033$ & $1105\pm42$ &  & $1.947\pm0.029$ & $3.5304\pm0.0044$ & $1076.2\pm5.3$ \\
G1 &  $0.95^{+0.12}_{-0.14}$ & $3.752^{+0.039}_{-0.046}$ & $846\pm38$ &  &  $1.052\pm0.022$ & $3.7179\pm0.0066$ & $876.3\pm6.2$ \\
G2 &  $2.51^{+0.18}_{-0.21}$ & $3.456\pm0.023$ & $1171\pm31$ &  & $1.794\pm0.025$ & $3.5548\pm0.0042$ & $1047.2\pm4.9$ \\
G3 &  $2.66\pm0.16$ & $3.440\pm0.018$ & $1193\pm25$ &  & $1.392\pm0.016$ & $3.6313\pm0.0035$ & $962.3\pm3.7$ \\
Total$^1$ & $12.40^{+0.29}_{-0.34}$ & - & - & & $8.268\pm0.016$ & - & - \\
\enddata
\tablecomments{$^1$The total mass is computed by summation of the virial mass from 5 fitted halos. $^2$For the WL only result, we use the reduced shear to fit NFW profiles. In case of the WL + SL, we use the 2-dimensional $\kappa$ distributions to fit NFW profiles (see \textsection\ref{NFW_fitting} for more details).}
\end{deluxetable*}

\begin{deluxetable*}{cccccccc}\label{NFW_fitting_result} 
\tablecaption{Comparison of the WL NFW fitting results with the main mass model}
\tablehead {
\colhead{} &
\colhead{} &
\colhead{WL} &
\colhead{} &
\colhead{} & 
\colhead{} &
\colhead{WL + SL} &
\colhead{}
\\ \cline{2-4}  \cline{6-8} 
\colhead{Component} &
\colhead{$r_s$} &
\colhead{$c_{200}$} &
\colhead{$M_{200}$} &
\colhead{} & 
\colhead{$r_s$} &
\colhead{$c_{200}$} &
\colhead{$M_{200}$}
\\
\colhead{} &
\colhead{(kpc)} & 
\colhead{} &
\colhead{($10^{14}M_{\odot}$)} &
\colhead{} &
\colhead{(kpc)} &
\colhead{} &
\colhead{($10^{14}M_{\odot}$)}
}
\startdata
BCG-North &  $521^{+80}_{-100}$ & $2.87^{+0.30}_{-0.38}$ & $4.96^{+0.89}_{-1.2}$ &  & $116.71^{+0.51}_{-0.74}$ & $9.996^{+0.043}_{-0.020}$ & $2.451\pm0.028$ \\
BCG-South &  $349^{+60}_{-100}$ & $3.41^{+0.50}_{-0.66}$ & $2.37^{+0.46}_{-0.77}$ &  & $106.27^{+0.68}_{-1.2}$ & $9.915^{+0.085}_{-0.037}$ & $1.827\pm0.024$ \\
G1 &  $222^{+40}_{-100}$ & $3.62^{+0.88}_{-1.0}$ & $0.63^{+0.11}_{-0.25}$ &  &  $104.4^{+1.8}_{-3.7}$ & $8.01^{+0.22}_{-0.14}$ & $0.912^{+0.017}_{-0.021}$ \\
G2 &  $637^{+90}_{-100}$ & $2.25^{+0.22}_{-0.27}$ & $4.41^{+0.71}_{-0.80}$ &  & $134.4\pm2.3$ & $8.13\pm0.12$ & $2.038\pm0.024$ \\
G3 &  $786\pm100$ & $1.89^{+0.15}_{-0.24}$ & $4.92\pm0.60$ &  & $179.8\pm2.8$ & $6.025\pm0.083$ & $1.985\pm0.018$ \\
Total$^1$ & - & - & $17.3\pm1.0$ & & - & - & $9.214\pm0.017$ \\
\enddata
\tablecomments{$^1$The total mass is computed by summation of the virial mass from 5 fitted halos. $^2$For the WL only result, we use the reduced shear to fit NFW profiles. In case of the WL + SL, we use the 2-dimensional $\kappa$ distributions to fit NFW profiles (see \textsection\ref{NFW_fitting} for more details).}
\end{deluxetable*}

\begin{figure*}
\centering
\includegraphics[width=\textwidth]{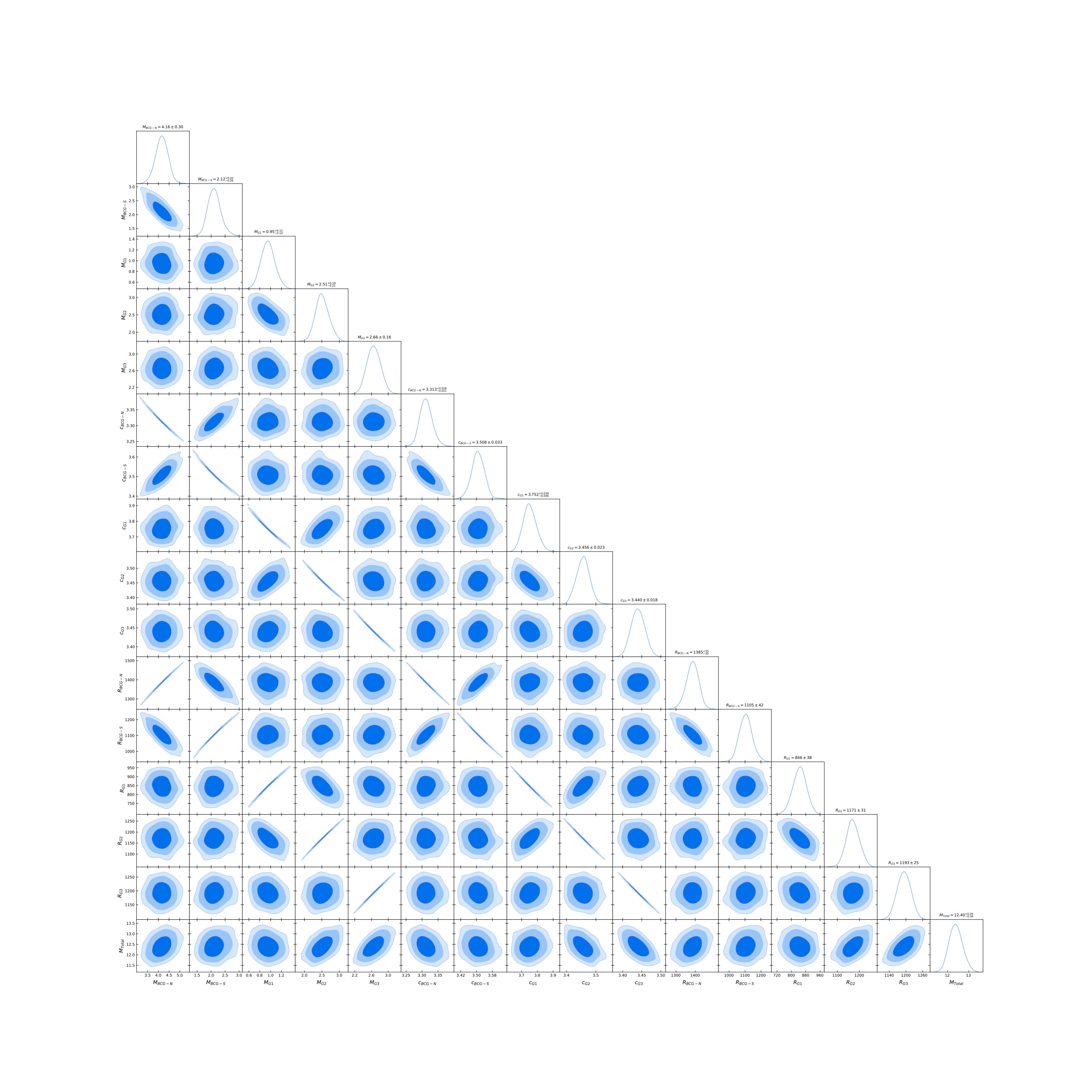}
\caption{WL-only NFW fitted results with $M-c$ relation of \citet{2008MNRAS.390L..64D}. Unit of mass and virial radius are $10^{14}M_{\odot}$ and kpc, respectively.}
\label{WL_mcmc_Mc}
\end{figure*}

\begin{figure*}
\centering
\includegraphics[width=\textwidth]{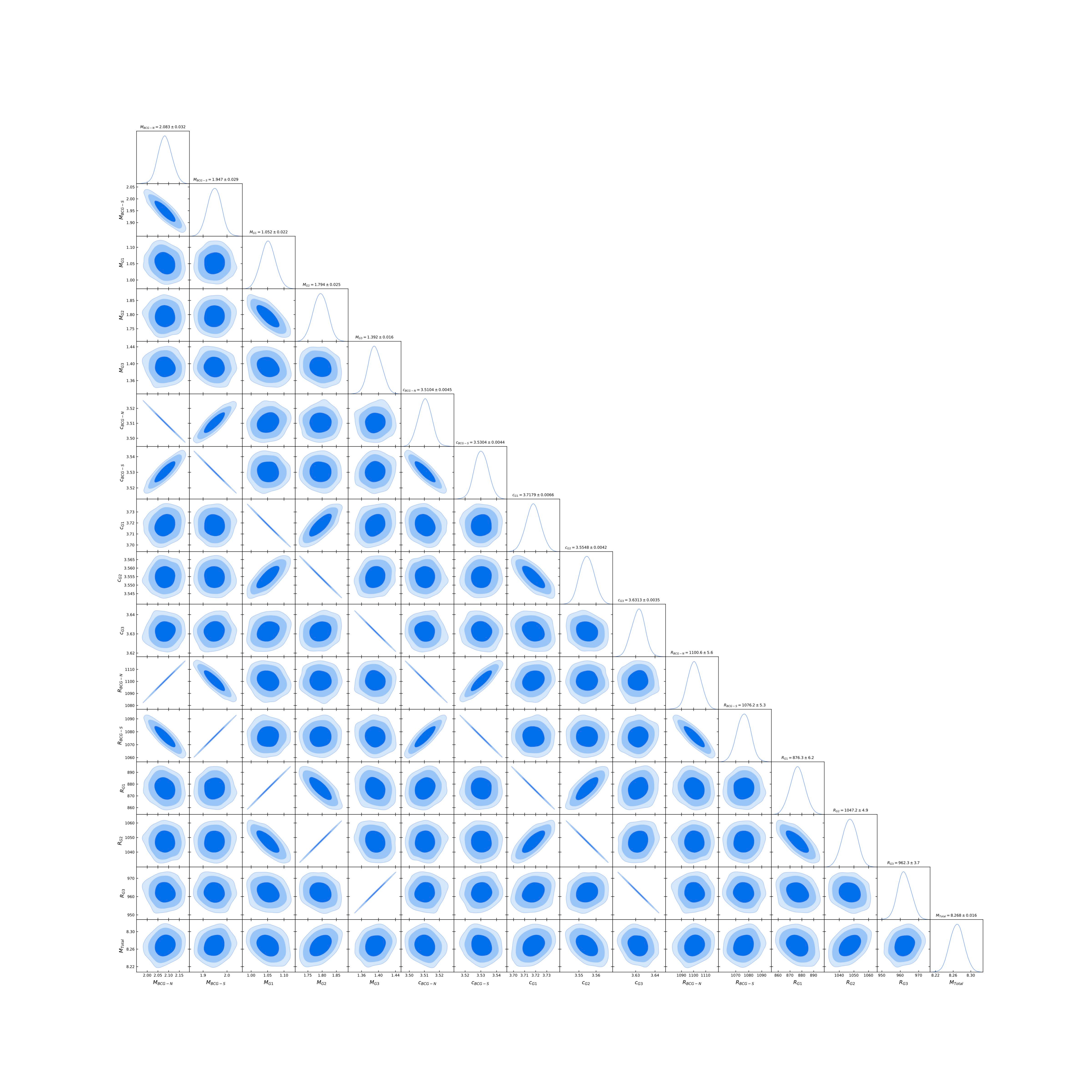}
\caption{WL + SL NFW fitted results with $M-c$ relation of \citet{2008MNRAS.390L..64D}. Unit of mass and virial radius are same as Figure~\ref{WL_mcmc_Mc}.}
\label{WL_SL_mcmc_Mc}
\end{figure*}

\begin{figure*}
\centering
\includegraphics[width=\textwidth]{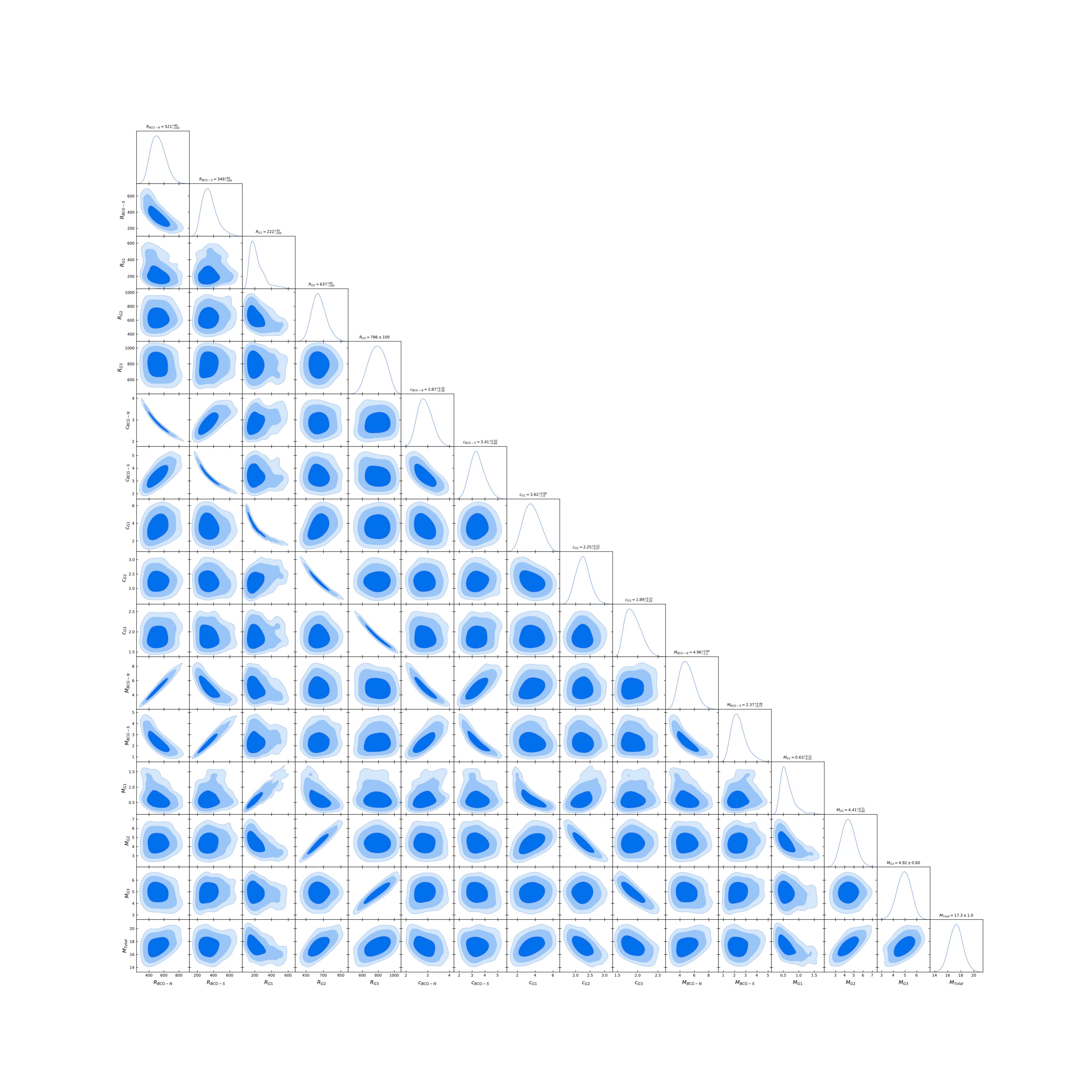}
\caption{WL-only NFW fitted results without $M-c$ relation. Unit of mass and scale radius are $10^{14}M_{\odot}$ and kpc, respectively.}
\label{WL_mcmc}
\end{figure*}

\begin{figure*}
\centering
\includegraphics[width=\textwidth]{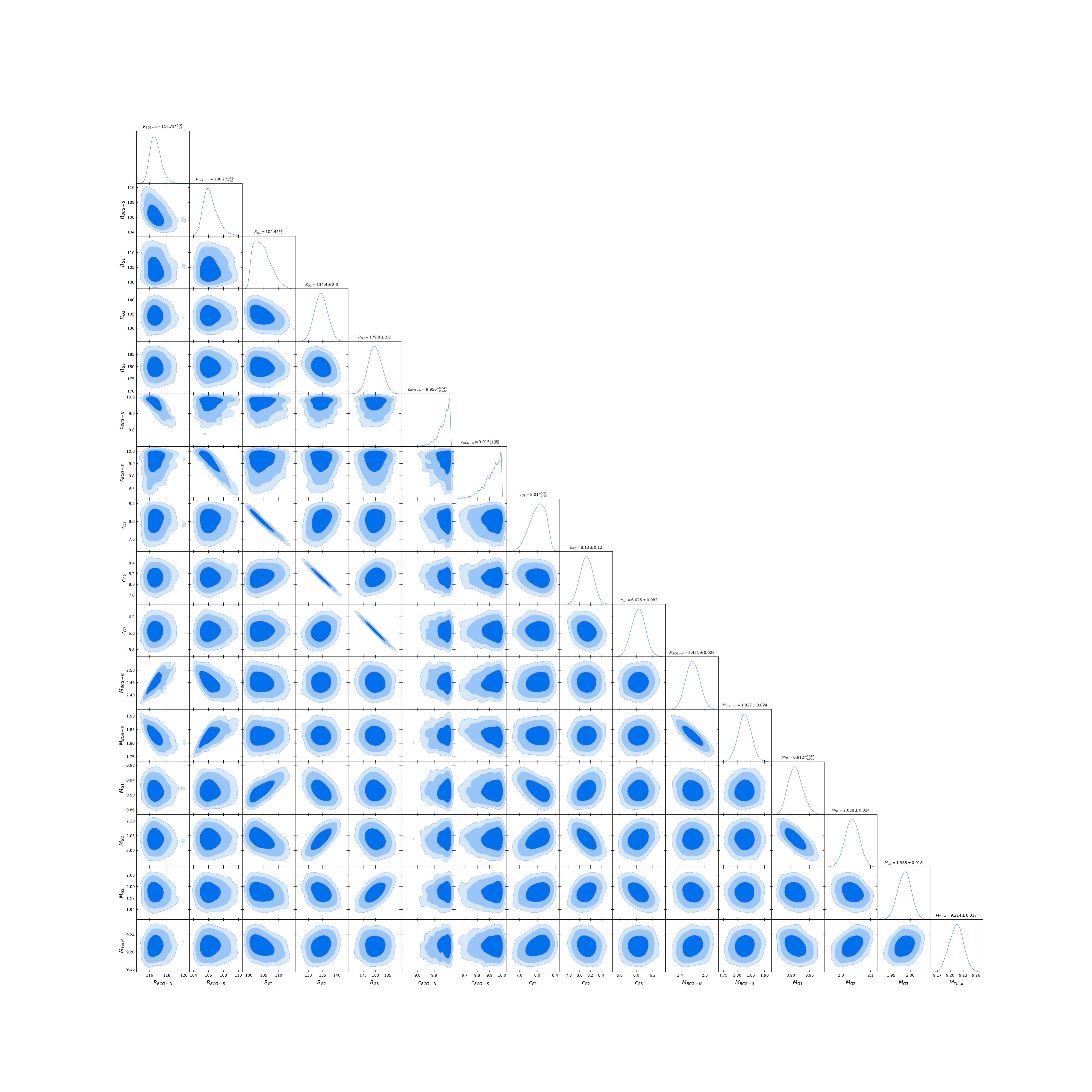}
\caption{WL + SL NFW fitted results without $M-c$ relation. Unit of mass and scale radius are same as Figure~\ref{WL_mcmc}.}
\label{WL_SL_mcmc}
\end{figure*}

\begin{figure*}
\centering
\includegraphics[width=0.7\textwidth]{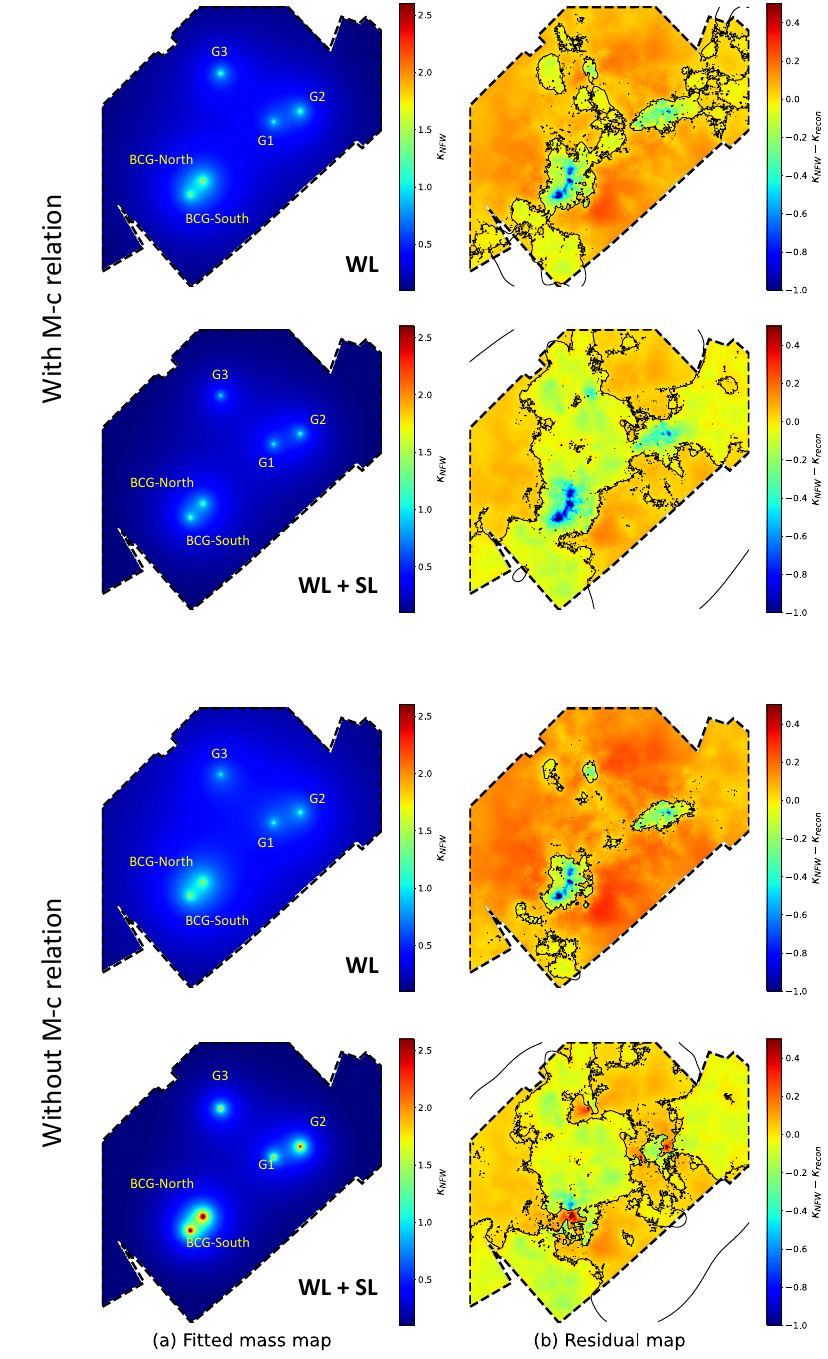} 
\caption{Residual mass maps between the NFW fitting and the main WL+SL mass maps. (a) Two-dimensional mass models created from the best-fit NFW parameters. (b) Subtraction of the main WL+SL mass map from the models shown in (a). The first and second (third and fourth) rows display the results obtained with (without) the $M-c$ relation of \citet{2008MNRAS.390L..64D}.
The black dashed lines indicate the footprints of the JWST observations. The black solid lines indicate regions where the residual is zero. See text for discussions.}
\label{NFW_fitting_2d}
\end{figure*}

The cumulative mass profiles that we present in \textsection\ref{mass_measurement} cannot be
used to isolate the mass properties of individual substructures because the measured convergence is the
result of the superposition of multiple halo profiles. Here, we determine the substructure properties by simultaneously fitting five NFW profiles.
A2744 is a complex system with a number of merging features including sophisticated X-ray morphologies and radio relics. Hence, this NFW fitting is not expected to produce results devoid of bias,
as mergers are likely to cause substantial deviations from NFW descriptions in cluster mass profiles
\citep[e.g.,][]{2023ApJ...945...71L}. Nevertheless, simultaneous multi-halo fitting can significantly diminish the impact of neighboring substructures. 

We perform simultaneous NFW fitting in two approaches. The first method is to fit five NFW profiles to our WL shape catalogs. In the second approach, we fit five NFW profiles to the convergence map obtained from the WL+SL mass reconstruction. In Tables~\ref{NFW_fitting_result_Mc} and \ref{NFW_fitting_result}, we display the results with and without the $M-c$ relation of \citet{2008MNRAS.390L..64D}, respectively. We also present the posterior distributions of all free parameters in Figures~\ref{WL_mcmc_Mc},~\ref{WL_SL_mcmc_Mc},~\ref{WL_mcmc}, and ~\ref{WL_SL_mcmc}. 
When we assume the $M-c$ relation of \citet{2008MNRAS.390L..64D}, both the WL and WL+SL results show
similar concentration values. However, in the results obtained without the $M-c$ relation, the WL+SL model gives considerably larger concentration values, which are attributed to the availability of the SL constraints near the mass peaks. The merging cluster simulations of \citet{2023ApJ...945...71L} demonstrate that the concentrations tend to increase because mass infalls occur in post-collision systems. However, since WL does not densely sample the signals near the halo centers, the concentrations are biased low, which in turn leads to an overestimation of the cluster mass in WL analysis. The extent of this overestimation depends on the state of the merger, with the factor potentially reaching as high as $2-3$. The systematic differences in both concentration and mass shown in Table~\ref{NFW_fitting_result} are consistent with the predictions of \citet{2023ApJ...945...71L}. 

In order to visualize the two-dimensional difference between the NFW fitting results and the main WL+SL mass map, we present the residual (subtraction of the WL+SL mass map from the NFW fitting result) mass maps in Figure~\ref{NFW_fitting_2d}.
Although the details differ between the two results with and
without the $M-c$ relations, similar trends are present in the residual maps.
The shear-based models produced with the best-fit NFW parameters (first and third rows) yield lower mass densities around the five mass peaks while they predict higher values in the outskirts. This illustrates that if we attempt to estimate the total mass of A2744 using the extrapolation of the shear-based NFW fitting results, the procedure will result in severe overestimation.
The residuals created with the convergence-based models (second and fourth rows) are somewhat more complex and show larger azimuthal variations.

\bibliographystyle{apj}
\bibliography{main}

\end{document}